\documentclass[reprint,twocolumn,aps,showpacs,amsmath,amssymb,prb,superscriptaddress]{revtex4-1}
\usepackage{color}
\usepackage{graphicx}
\usepackage{dcolumn}
\usepackage{bm}

\begin{document}

\title{Phase separation in the non-equilibrium Verwey transition in magnetite}

\author{F. Randi$^{**}$}%
\affiliation{Department of Physics, Universit\`{a} degli Studi di Trieste, 34127 Trieste, Italy}

\author{I. Vergara$^{**}$}%
\affiliation{II. Physikalisches Institut, Universit\"{a}t zu K\"{o}ln, 50937 K\"{o}ln, Germany}%

\author{F. Novelli}%
\affiliation{Sincrotrone Trieste SCpA, 34127 Basovizza, Italy}%

\author{M. Esposito}%
\affiliation{Department of Physics, Universit\`{a} degli Studi di Trieste, 34127 Trieste, Italy}

\author{M. Dell'Angela}%
\affiliation{Sincrotrone Trieste SCpA, 34127 Basovizza, Italy}%

\author{V. A. M. Brabers}%
\affiliation{Department of Physics, Eindhoven University of Technology, Eindhoven, The Netherlands}


\author{P. Metcalf}%
\affiliation{Purdue University, School of Materials Engineering, West Lafayette, Indiana 47907, USA}%

\author{R. Kukreja}%
\affiliation{Stanford Institute for Energy and Materials Sciences, SLAC National Accelerator Laboratory, 2575 Sand Hill Road, Menlo Park, California 94025, USA}%

\author{H. A. D\"{u}rr}%
\affiliation{Stanford Institute for Energy and Materials Sciences, SLAC National Accelerator Laboratory, 2575 Sand Hill Road, Menlo Park, California 94025, USA}%

\author{D. Fausti}%
\email[Corresponding author: ]{daniele.fausti@elettra.eu}
\affiliation{Department of Physics, Universit\`{a} degli Studi di Trieste, 34127 Trieste, Italy}
\affiliation{Sincrotrone Trieste SCpA, 34127 Basovizza, Italy}%

\author{M. Gr\"{u}ninger}
\affiliation{II. Physikalisches Institut, Universit\"{a}t zu K\"{o}ln, 50937 K\"{o}ln, Germany}%

\author{F. Parmigiani}%
\affiliation{Department of Physics, Universit\`{a} degli Studi di Trieste, 34127 Trieste, Italy}%
\affiliation{Sincrotrone Trieste SCpA, 34127 Basovizza, Italy}%
\affiliation{International Faculty, Universit\"{a}t zu K\"{o}ln, 50937 K\"{o}ln, Germany}

\date{\today}

\begin{abstract}
We present equilibrium and out-of-equilibrium studies of the Verwey transition in magnetite. In the equilibrium optical conductivity, we find a step-like change at the phase transition for photon energies below about 2\,eV.\@ The possibility of triggering a non-equilibrium transient metallic state in insulating magnetite by photo excitation was recently demonstrated by an x-ray study. Here we report a full characterization of the optical properties in the visible frequency range across the non-equilibrium phase transition. Our analysis of the spectral features is based on a detailed description of the equilibrium properties. The out-of-equilibrium optical data bear the initial electronic response associated to localized photo-excitation, the occurrence of phase separation, and the transition to a transient metallic phase for excitation density larger than a critical value. This allows us to identify the electronic nature of the transient state, to unveil the phase transition dynamics, and to study the consequences of phase separation on the reflectivity, suggesting a spectroscopic feature that may be generally linked to out-of-equilibrium phase separation.
\end{abstract}

\maketitle

\section{Introduction}

\begin{figure}[t]
\centering
\includegraphics[width=\linewidth]{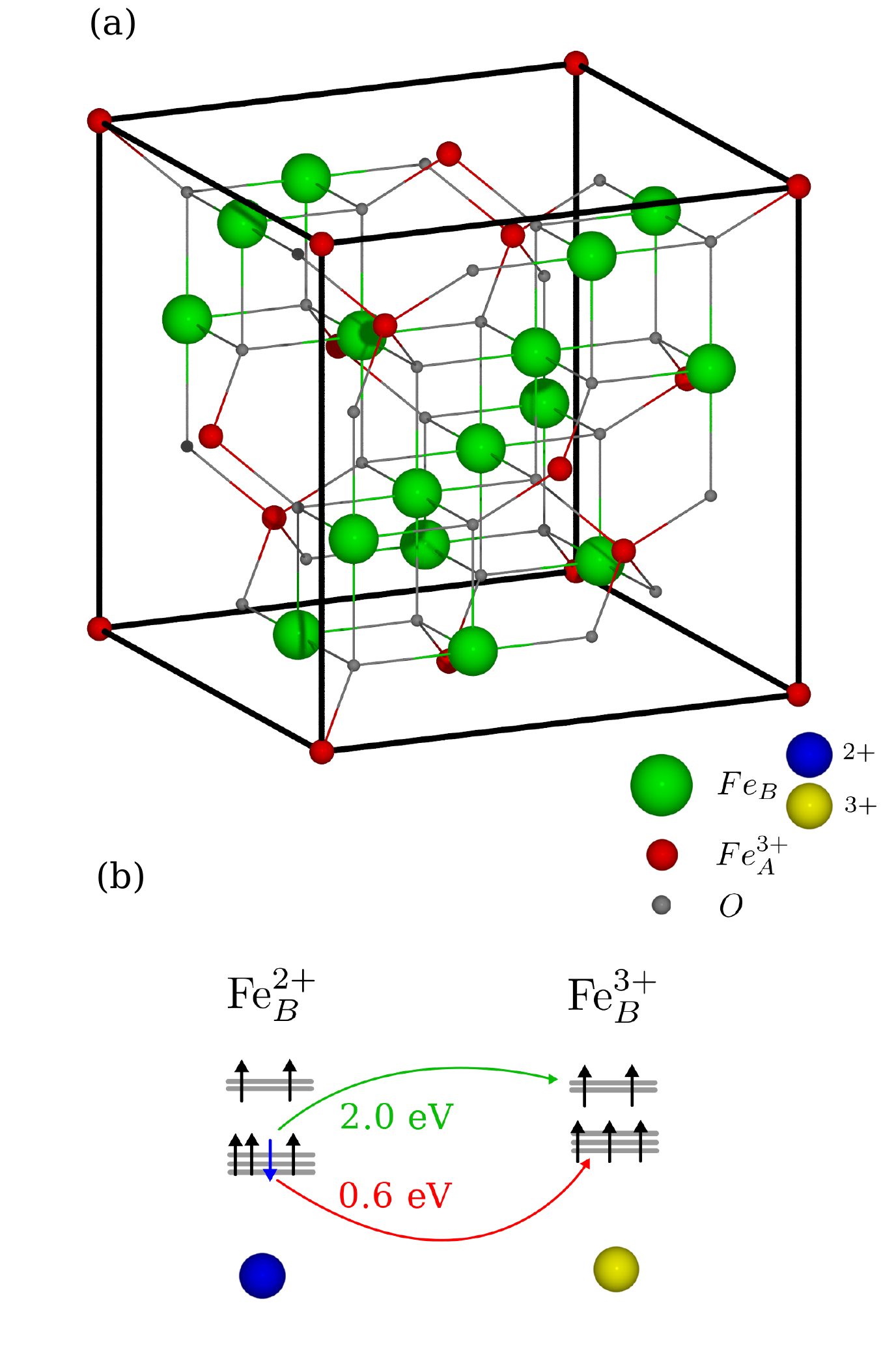} 
\caption{(a) Inverse spinel high temperature structure of magnetite.
(b) Sketch of the Fe$_B^{2+}\rightarrow$Fe$_B^{3+}$ transitions.
}
\label{fig:StructureTransitions}
\end{figure}

The Verwey phase transition, occuring at $T_V$\,=\,123\,K in magnetite (Fe$_3$O$_4$), is presumably the most studied metal-insulator transition in the large family of transition-metal oxides.\cite{Verwey1939,Khomskii2014}
Detailed investigations of the structure revealed a stunning complexity.\cite{Senn2012,Nazarenko2006,Lorenzo2008,Garcia2009,Weng2012,Blasco2011,Huang2006,Schlappa2008,Tanaka2012,Wright2002,Fuji1975,Shapiro1976}
Magnetite crystallizes with an inverse spinel structure (figure~\ref{fig:StructureTransitions}a), characterized by two different groups (A and B) of Fe ion sites.
Group A is constituted by tetrahedrally coordinated Fe$_A^{3+}$ sites with a $3d^5$ electronic configuration with spin $S$=5/2. Group B, instead, is made of octahedrally coordinated sites formally occupied by Fe$_B^{3+}$ ($3d^5$, $S$=5/2) and Fe$_B^{2+}$ ($3d^6$, $S$=2) ions. At $T_c=858$\,K, ferrimagnetic order sets in with ferromagnetic coupling within the B sublattice and antiferromagnetic coupling between A and B sites such that the A sites carry minority spins $\downarrow$. The original picture of the Verwey transition at $T_V$ involves a metallic high-temperature phase showing an equal number of Fe$_B^{3+}$ and Fe$_B^{2+}$ ions randomly distributed on the B sites opposed to an insulating low-temperature phase with a charge-ordered B sublattice~\cite{Verwey1939}, breaking the cubic symmetry in favour of a monoclinic one. Since the B ions form a frustrated pyrochlore lattice, the charge-order superstructure is rather involved, reflecting the competition between Coulomb interactions and the coupling to both the lattice and the orbital degrees of freedom.\cite{Wright2002,Khomskii2014} Recently, it has been shown that the picture may need to be refined even further.\cite{Senn2012} On top of the charge order, x-ray studies suggest that the single minority $\downarrow$ electron of an Fe$_B^{2+}$ ion is delocalized over the neighbouring Fe$_B^{3+}$ sites, forming linear units of three Fe sites dubbed trimerons.\cite{Senn2012} Such units are organized in a network where different trimerons are connected via angles of 60$^\circ$ or 120$^\circ$. Since the Fe$_B^{3+}$ sites are part of up to three different trimerons (sharing different $t_{2g}$ orbitals\cite{Senn2012}), the trimeron lattice tends to equalize the charges on the Fe$_B$ sites and to increase the polarizability.\cite{vandenBrink2008}
Furthermore, having the minority spin delocalized on different sites reduces the expected entropy associated to the Verwey transition to values closer to experimentally observed ones.\cite{Sheperd1985,Senn2012} In this interpretation, it has been proposed~\cite{Senn2012} that the Verwey transition should be seen as a transition between a frozen trimeron network and a fluctuating network with shorter correlation length.

With x-ray pump-probe experiments\cite{deJong2013} some of us investigated the quench of the insulating phase after the excitation with ultrashort pulses at 1.55\,eV.\@
In that work it has been found that the light-driven structural change occurs in a two-step process: after the initial destruction of individual trimerons triggered by photo excitation, phase separation occurs yielding metallic and residual insulating regions.

Optical measurements provide an efficient tool to unravel the changes of electronic and structural properties at the phase transition.~\cite{Cavalleri2006,Chollet2005,Fausti2009,Baldini2015} In magnetite, the equilibrium optical conductivity\cite{Schlegel1979,Gasparov2000,Park1998} shows a broad Drude peak in the metallic phase as well as its suppression at the metal-insulator transition. The optical properties at higher energies are dominated by two features peaking at about 0.6\,eV and 2\,eV.\@ By comparison with LSDA+U results,\cite{Leonov2006} these features were attributed to excitations of the minority $\downarrow$ electrons from the Fe$_B^{2+}$ t$_{2g}$ levels to the Fe$_B^{3+}$ t$_{2g}$ and $e_g$ levels, respectively (see figure~\ref{fig:StructureTransitions}b). However, alternative interpretations invoking the A sites were proposed for the feature at 2\,eV.\cite{Park1998,Kim2007} Thus far, the behavior of these peaks at the Verwey transition has only been addressed at a qualitative level.

Here we report on detailed equilibrium and out-of-equilibrium measurements of the optical properties of magnetite in a broad spectral range. Ellipsometric data allow us to quantify the change of the equilibrium optical properties at the Verwey transition. The out-of-equilibrium measurements were performed under the same excitation conditions as used in combination with an x-ray probe in Ref.\ [\onlinecite{deJong2013}].
These measurements were performed at different temperatures to identify the analogies of the out-of-equilibrium insulator-to-metal transition with the thermodynamic one.
While confirming the already described overall phase-transition dynamics, we discuss the consequences of out-of-equilibrium phase separation on the transient reflectivity.
From these observations we propose a simple analysis to identify a spectroscopic feature that may be generally linked to out-of-equilibrium phase separation. Based on the analysis of the equilibrium data, we finally study the dynamics of the spectral features of the photo-induced transient state.

\section{Experiments}

We performed pump-probe measurements on magnetite using 1.55\,eV-centred 80\,fs pulses as pumps and broadband pulses with spectral components extending
from 1.7\,eV to 2.5\,eV as probes. The laser repetition rate was 250\,kHz. More details of the set-up have been described by Novelli \textit{et al.} in Ref.\ [\onlinecite{Novelli2012}]. No physical correction of the chirp of the broadband white light pulses was performed, but a post-processing correction the chirp was applied to the data. The experiment was performed with the sample at 35\,K, 80\,K, and 140\,K, the latter one being above the Verwey transition temperature. The out-of-equilibrium results reported here have been measured on two different samples, grown by the floating-zone technique in independent laboratories.~\cite{deJong2013,Kuipers1976} One of the samples was the same as used by de Jong \textit{et al.}\cite{deJong2013} in the time-resolved x-ray diffraction experiment. For the measurements of the equilibrium properties, we used a sample oriented in the [100] direction and polished to obtain an optically smooth surface. In the energy range from 0.75\,eV to 3.5\,eV, ellipsometric data were acquired with a rotating-analyzer ellipsometer (Woollam VASE) equipped with a retarder between polarizer and sample. The angle of incidence was $70^\circ$. The sample was mounted in a liquid-He flow cryostat with optical windows under UHV conditions ($<10^{-9}$\, mbar). For the analysis of the ellipsometric data, we assumed cubic symmetry and considered a surface roughness of 4\,nm. The analysis directly yields the complex dielectric function $\varepsilon(\omega)$\,=\,$\varepsilon_1(\omega) + i \varepsilon_2(\omega)$, or, equivalently, the complex optical conductivity $\sigma(\omega) \propto i [\varepsilon(\omega)-1]$.

\section{Results}

\subsection{Equilibrium optical properties}
\label{subsec:Equilibrium}

\begin{figure}[t]
\centering
\includegraphics[width=0.95\linewidth]{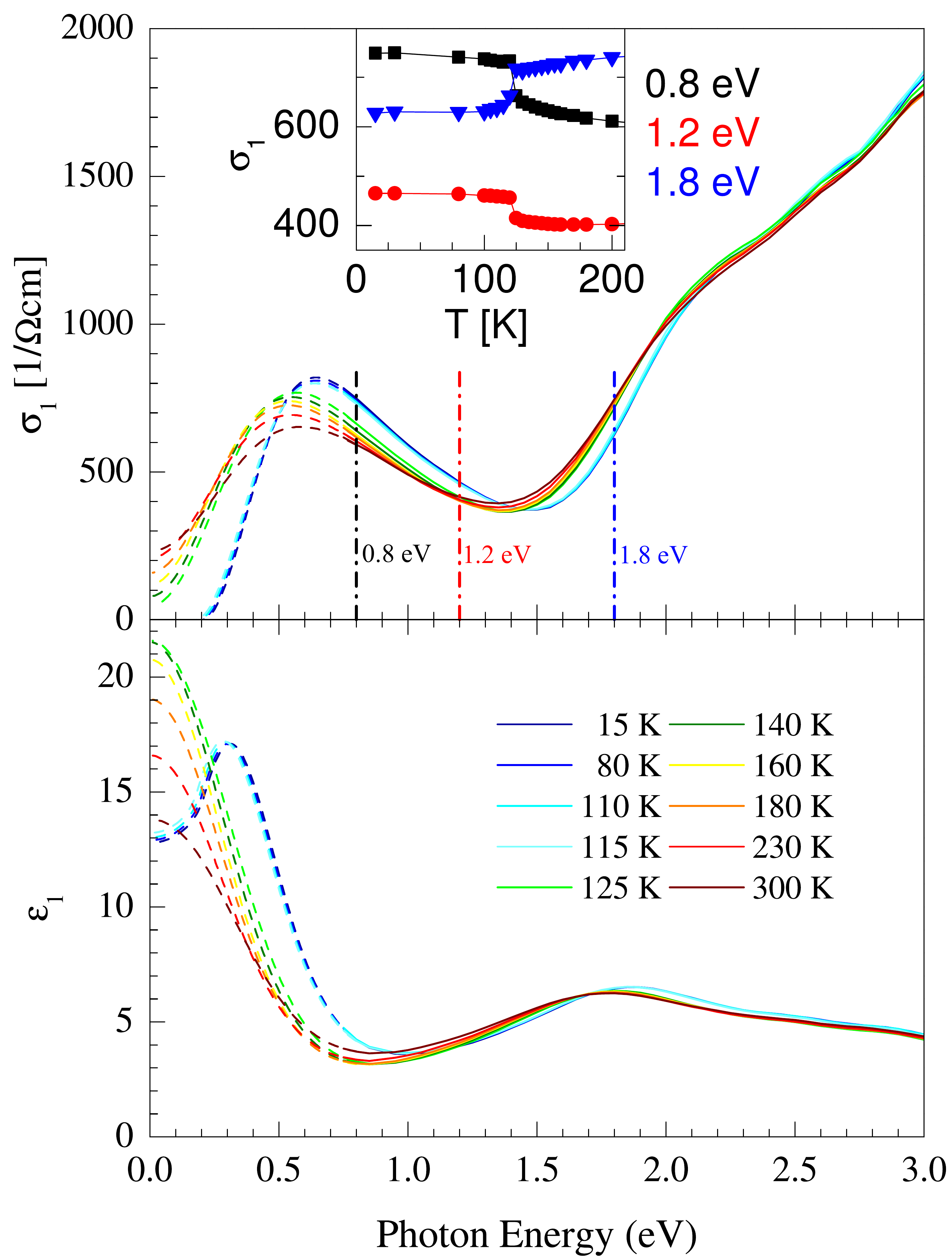}
\caption{Real parts $\sigma_1(\omega)$ and $\varepsilon_1(\omega)$ of the optical conductivity (top) and the dielectric function (bottom)
as determined by ellipsometry (solid lines). Dashed lines show extrapolations based on fits of the ellipsometric data (see main text). The opening of a gap in $\sigma_1$ at low temperatures gives rise to the peak in $\epsilon_1$ at about 0.3 eV. Inset: The temperature dependence of $\sigma_1$ at the three frequencies indicated in the main panel highlights step-like changes at $T_V$.
}
\label{fig:OpticalEquilibrium}
\end{figure}

\begin{figure}[t]
\centering
\includegraphics[width=0.95\linewidth]{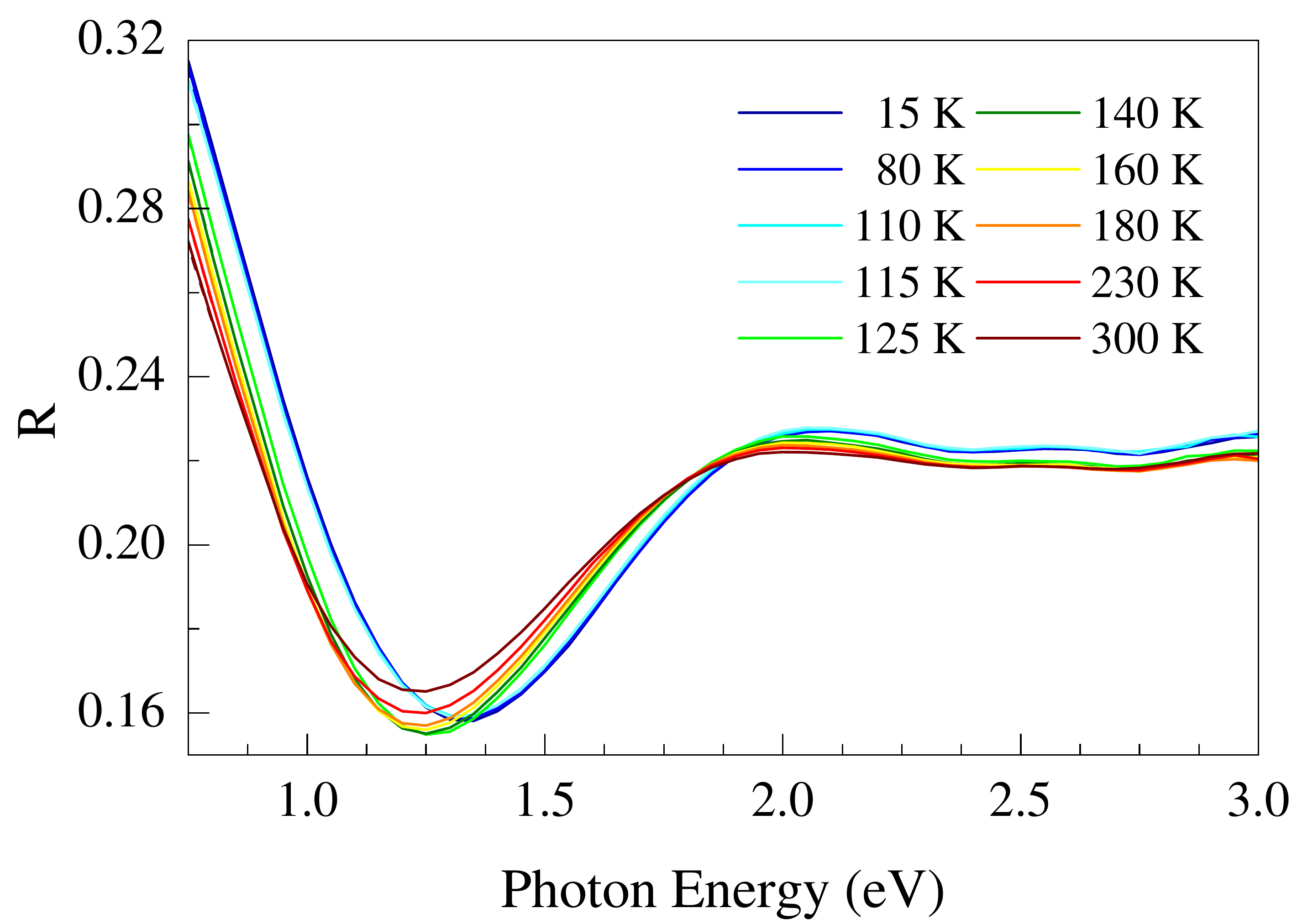}
\caption{Equilibrium reflectivity derived from the ellipsometric results plotted in figure~\ref{fig:OpticalEquilibrium}.
}
\label{fig:refl}
\end{figure}

The solid lines in figure~\ref{fig:OpticalEquilibrium} depict $\sigma_1(\omega)$ and $\varepsilon_1(\omega)$ as determined from the ellipsometric data at selected temperatures. Overall, the data agree with previous results which were based on a Kramers-Kronig analysis of reflectivity data.\cite{Schlegel1979,Gasparov2000,Park1998}
In $\sigma_1(\omega)$, strong absorption related to charge-transfer excitations from O$_{\text{2p}}$ states to Fe$_{\text{3d}}$ states sets in at about 2.5\,eV, while the two absorption bands peaking at about 0.6\,eV and 2\,eV were attributed to excitations within the Fe$_{\text{3d}}$ states.\cite{Gasparov2000,Park1998,Leonov2006,Kim2007} With increasing temperature, the spectral weight of the 0.6\,eV feature decreases, in agreement with the Kramers-Kronig results.\cite{Gasparov2000,Park1998} However, the literature data\cite{Gasparov2000,Park1998} do not address the precise behavior at the Verwey transition, and they disagree on the temperature dependence above 1.5\,eV.\@ This range is of particular importance for our pump-probe measurements with broadband probe pulses extending from 1.7\,eV to 2.5\,eV.\@ Ellipsometry is a self-normalizing technique which directly yields the complex dielectric function without the need to invoke a Kramers-Kronig analysis. It therefore is particularly well suited to determine the precise temperature dependence of the optical properties.\cite{Goessling2008a,Goessling2008b,Reul2012} The inset of figure~\ref{fig:OpticalEquilibrium} focusses on $\sigma_1(T)$ at three selected frequencies, revealing a step-like change of $\sigma_1(T)$ at $T_V$. This is a clear signature of the change of the electronic structure.

To provide a solid starting point for the analysis of the out-of-equilibrium optical data, we fitted the ellipsometric data using an oscillator model. The dashed lines in figure~\ref{fig:OpticalEquilibrium} show the extension of the optical properties to lower photon-energies provided by the model. We employed a Gaussian line shape for the features above 1.5\,eV, while a Tauc-Lorentz profile was assumed for the absorption band peaking at about 0.6\,eV since this band shows a gap-like feature in the insulating phase. In the metallic phase, the model includes a Drude peak describing free carriers, $\sigma_1^{\rm Drude}$\,=\,$\sigma_{\rm DC}/(1+\omega^2\tau^2)$. This Drude peak mainly contributes below the lower limit of our frequency range. Therefore, we fixed the two parameters $\sigma_{\rm DC}$ and $\tau$ of the Drude peak by using the measured DC resistivity and a temperature-independent value for the scattering rate $1/\tau$ which was adapted to describe the room-temperature data of Park and collaborators.\cite{Park1998} Consideration of the temperature dependence of this Drude peak via $\sigma_{\rm DC}(T)$ provides a more reliable determination of the properties of the prominent peak at 0.6\,eV, of which our data covers only the high-energy side (see supplemental materials~\cite{Supplemental}). Based on the assumption that this peak shows a Tauc-Lorentz profile we still may obtain a reasonable estimate of its properties as ellipsometry yields both $\varepsilon_1(\omega)$ and $\varepsilon_2(\omega)$, providing a strong constraint for the fits. This is corroborated by the reasonable agreement between our fits and the low-energy results for $\varepsilon_1$ of Ref.\ [\onlinecite{Schlegel1979}] and for $\sigma_1(\omega)$ of Refs.\ [\onlinecite{Gasparov2000}] and [\onlinecite{Park1998}] concerning, e.g., the peak frequency of about 0.6\,eV, the peak height, and the size of the gap $\Delta$ in the insulating phase. Our fits yields $\Delta$\,=\,0.2\,eV at 15\,K, which coincides with the value reported by Gasparov \textit{et al.}\cite{Gasparov2000} while Park \textit{et al.}\cite{Park1998} find 0.14\,eV.\@ Even though the quantitative results for the low-energy oscillator have to be taken with some care, our model provides an excellent basis for the analysis of the out-of-equilibrium data measured between 1.7\,eV and 2.5\,eV, well within the range covered by our ellipsometric data. The corresponding equilibrium reflectivity spectra are plotted in figure~\ref{fig:refl} at selected temperatures.

\begin{figure}[t]
\centering
\includegraphics[width=0.9\linewidth]{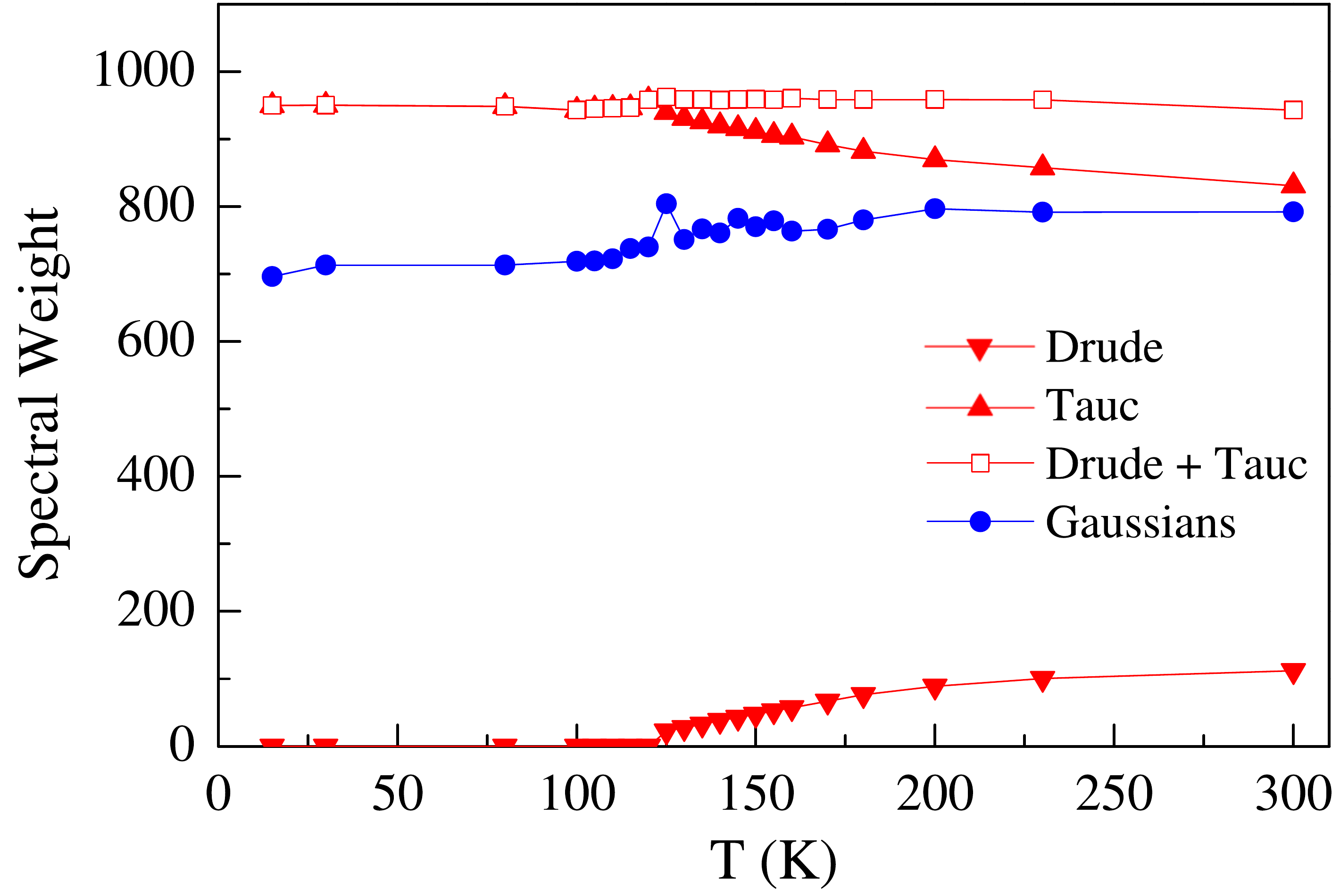}
\caption{Temperature dependence of the spectral weights of different oscillators used in the model (see dashed lines in figure~\ref{fig:OpticalEquilibrium}).
}
\label{fig:SW}
\end{figure}

The spectral weight denotes the integral of a given feature in $\sigma_1(\omega)$. The temperature dependence of the spectral weights of the different contributions to the oscillator model are depicted in figure~\ref{fig:SW}. Following the gradual change of the DC resistivity, the Drude peak loses spectral weight upon cooling and vanishes at the metal-insulator transition at $T_V$. This change of spectral weight is roughly compensated by the Tauc-Lorentz oscillator at 0.6\,eV, the sum of the spectral weights of the two features is nearly independent of temperature (see open symbols). This is consistent with the results of Gasparov \textit{et al.},\cite{Gasparov2000} who reported that the spectral weight is roughly independent of $T$ below about 0.8\,eV.\@ This agreement further supports the reliability of our model. The gradual change of both the Drude peak and the Tauc-Lorentz oscillator extends to temperatures far above $T_V$, which was attributed to short-range charge order.\cite{Park1998} Note that this gradual change can also be observed in the ellipsometry result for $\sigma_1(T)$ (see inset of figure~\ref{fig:OpticalEquilibrium}), i.e., independent of any oscillator model. Qualitatively, the $T$ dependence in the visible range (blue symbols in figure~\ref{fig:SW}) is similar to that of the Tauc-Lorentz oscillator, albeit with the opposite sign.

\subsection{Out-of-equilibrium optical properties}

\begin{figure}[t]
\centering
\includegraphics[width=\linewidth]{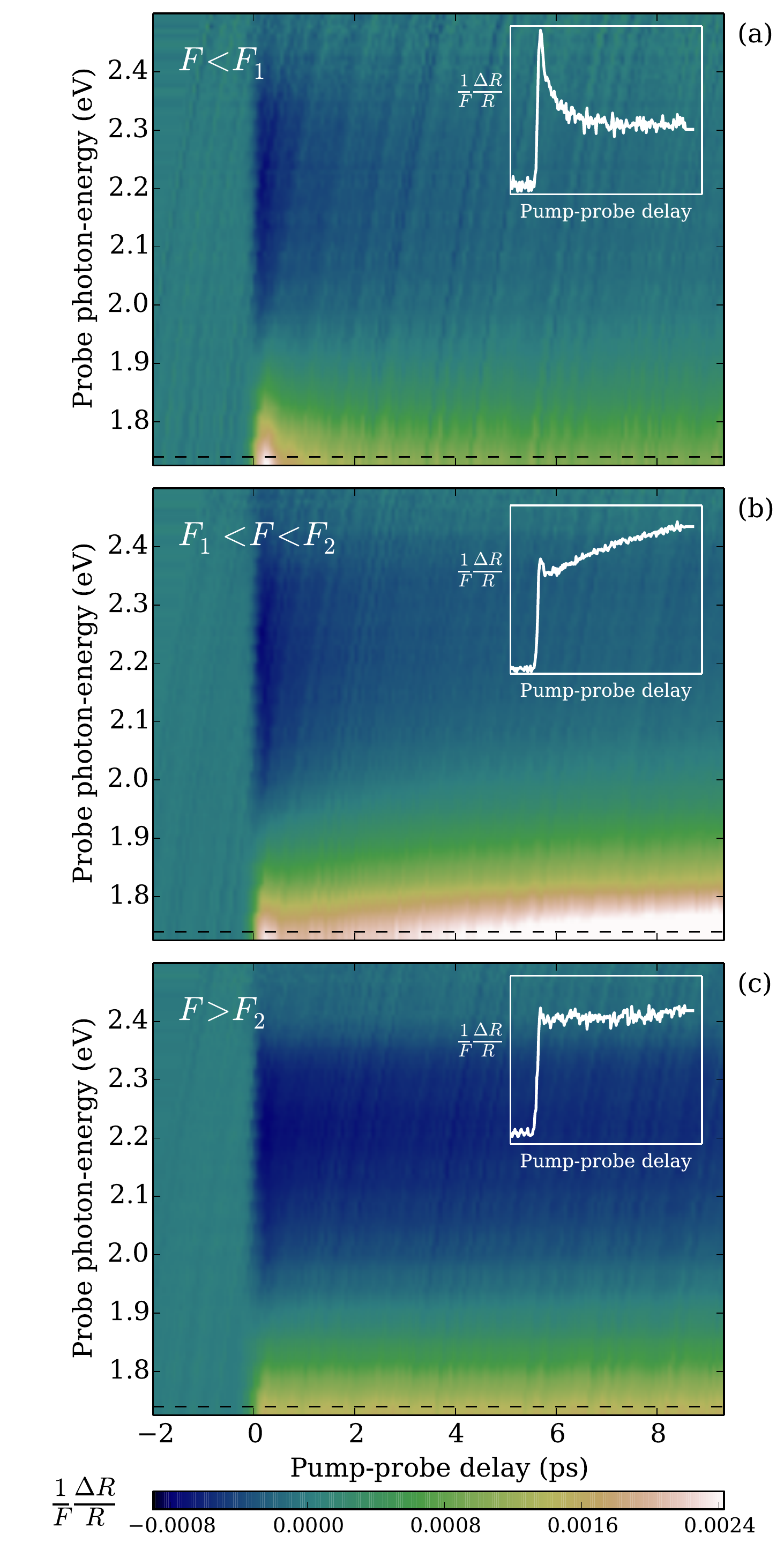}
\caption{
Normalized relative variation of the reflectivity $\frac{1}{F}\frac{\Delta R}{R}(h\nu, t)$ at 35 K for pump fluences F of (a) 0.5 mJ cm$^{-2}$, (b) 4.6 mJ cm$^{-2}$, and (c) 7.1 mJ cm$^{-2}$, characteristic of the three regimes of low, intermediate, and high fluence. Insets: $\frac{\Delta R}{R}(t)$ at 1.74 eV for the respective fluences. Dashed lines: photon-energy corresponding to the insets.
}
\label{fig:dRR}
\end{figure}

Our out-of-equilibrium data can be separated into two classes. The data measured at 35\,K and 80\,K (T\,$<$\,T$_V$) show similar behavior, which we will argue to be the out-of-equilibrium equivalent of the Verwey transition. The data measured at 140\,K (T\,$>$\,T$_V$), instead, do not show the distinctive features displayed below T$_V$, linked to the photo-induced phase transition. Since the results at 35\,K and 80\,K are almost completely equivalent, for the sake of clarity we will describe and discuss explicitly only the results at 35\,K, underlining where the differences with the data at 80\,K arise and how these differences support the picture drawn for 35 K (see the supplemental materials~\cite{Supplemental} for plots of the data at 80 K). The results at 35\,K allow us to identify three regimes of pump fluence in which, with T\,$<$\,T$_V$, the relative variation of the reflectivity ($\frac{\Delta R}{R}(t,h\nu)$) behaves qualitatively different. The only difference with the results at 80\,K are the pump fluences delimiting these intervals, which are shifted to lower values for higher temperature. In figure~\ref{fig:dRR} we plot three data sets, each of them representative of one of the three regimes.

Below a pump fluence of F$_1$ = 2.7 mJ cm$^{-2}$, the relative variation of the reflectivity $\frac{\Delta R}{R}(t,h\nu)$ behaves as shown in figure~\ref{fig:dRR}a. In this regime, which in the following we will call \textit{low fluence}, the response has two distinctive features. First of all, after a very fast increase at pump-probe delay $t$\,=\,0 at low probing energies around 1.8\,eV, $\frac{\Delta R}{R}$ decays exponentially with a characteristic time scale of 0.9 ps to a non-zero thermal plateau (see inset of figure~\ref{fig:dRR}a). The second feature is that $\frac{\Delta R}{R}$ scales linearly with the fluence. This behavior is typical of the creation of excitations, whose density scales linearly with the fluence. The excitations then decay bringing the system to a thermalized state with a temperature different from the initial one. The lifetime of this transient state is determined by the thermal conductivity of the system.

In the \textit{intermediate fluence} regime between F$_1$ and F$_2$ = 5.1 mJ cm$^{-2}$, the behavior of $\frac{\Delta R}{R}(t,h\nu)$ is the one plotted in figure~\ref{fig:dRR}b. Again, we can identify two characteristic features. First, the excitation is followed by two distinct dynamics. After quickly decaying for a very short time ($<$1 ps), the response grows again on a slower time scale $\tau_2$ (see inset of figure~\ref{fig:dRR}b). This happens more pronouncedly on the low-energy side of the probed range. Second, while the amplitude of the initial peak (t $\simeq 0.2$ ps) still scales linearly with the fluence (as in the low fluence regime), the amplitude of the long-time dynamics does not and its characteristic timescale is not constant with the fluence. From these considerations we can say that the creation of a sufficient number of excitations in the system triggers a new dynamical response, evolving on a larger time scale than the decay of the excitations.

In the \textit{high fluence} regime, starting from F$_2$, $\frac{\Delta R}{R}$ progressively loses the mentioned features, up to the point that for the highest measured fluence (7.1 mJ cm$^{-2}$) it behaves as a switch on the low-energy side of the spectrum, as shown in figure~\ref{fig:dRR}c. In this regime, $\frac{\Delta R}{R}$ is almost independent of the fluence. Moreover, apart from a small relaxation on the high-energy side of the spectrum, $\frac{\Delta R}{R}$ displays a step-like behavior at t = 0 and does not evolve anymore for times at least longer than 10\,ps.

\section{Discussion}
By means of time-resolved x-ray diffraction, de Jong et al.~\cite{deJong2013} have shown that holes in the charge-ordered lattice, purportedly the trimeronic lattice~\cite{Senn2012}, are produced upon excitation by the pump pulse. If the pump fluence is lower than F$_1$, the lattice thermalizes to a higher temperature, but retains the global symmetry of the low-temperature charge-ordered phase (\textit{low fluence} regime). If the fluence exceeds F$_1$, nucleation of volumes with the symmetry of the high-temperature phase is triggered. This leads to phase separation, i.e., coexistence of uncorrelated charge-ordered regions and metallic ones (\textit{intermediate fluence} regime). Our time-resolved spectroscopic data confirm this scenario. Moreover, we also explored higher fluences (F $>$ F$_2$), where the dynamics is different.

Our discussion will proceed as follows. First, we will discuss the intermediate fluence regime and we will show that the spectral feature of the long-time response corresponds to the nucleation of the high-temperature phase. We will afterwards identify the general consequences of phase separation on the separability of the dynamical properties of a system. We will then show that in the high fluence regime the system is, instead, immediately (i.e., on a timescale smaller than our experimental resolution) and homogeneously brought to the high-temperature phase and the nucleation process can no longer be observed in the out-of-equilibrium optical properties. Residual localized charge order is still present in the system, but the insulating region is progressively reduced as the fluence increases. This scenario emerges from the comparison of the out-of-equilibrium with the equilibrium measurements and from the study of the separability of the variation of the reflectivity as a function of time and probe-photon-energy.

The results obtained at 140 K confirm that our observations can be ascribed to a photo-induced phase transition. Above the Verwey transition temperature and for the measured fluences the response is linear and there is no evidence for a photo-induced phase transition. In the following we will use this substantial difference to benchmark part of the proposed analysis.

\subsection{Out-of-equilibrium phase transition and its relation to equilibrium}

In this section we first present plots of quantities parameterizing both the fast and the long-time response of the system and identify the critical fluences to initiate (F$_1$) and saturate (F$_2$) the out-of-equilibrium phase transition. Subsequently we relate the fluences F$_1$ and F$_2$ with effective temperatures of the system after the excitation and compare them with temperatures relevant in the equilibrium thermodynamics and optical properties.

\begin{figure}[t]
\centering
\includegraphics[width=\linewidth]{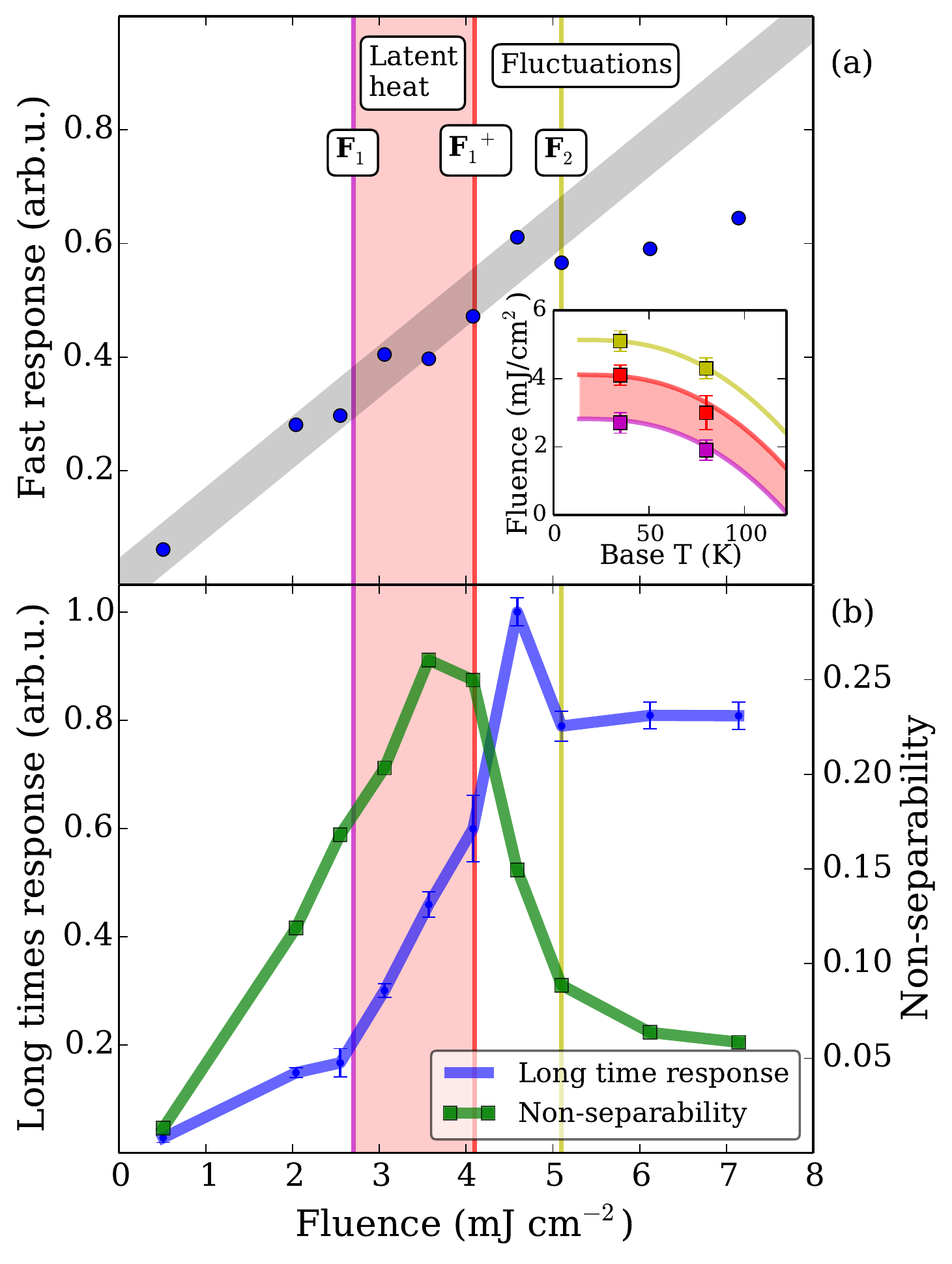}
\caption{
(a) ``Fast response'': $\frac{\Delta R}{R}$ at 1.74 eV photon-energy and 0.2 ps pump-probe delay.
(b) ``Long-time response'' (blue curve): $\frac{\Delta R}{R}$ at 1.74 eV photon-energy and 8 ps pump-probe delay. ``Non-separability'' (green curve): ratio of the second largest and largest singular values (see section~\ref{subsec:NonSeparability}).
(a inset) The squares are the characteristic fluences extracted from the out-of-equilibrium data as a function of the sample's temperature, corresponding to: onset of the non-linear response (magenta), beginning of the decreasing of the ``non-separability'' (red), saturation of the non-linear response (yellow). The lines represent equivalent fluences calculated from thermodynamic data needed to: reach T$_V^-$ (magenta), reach T$_V^+$ (red), reach 140 K (yellow). The red shaded area corresponds to fluences bringing the sample to T$_V^-$ and supplying part of the latent heat.
Vertical lines and shaded areas in the main figure mimic the inset.
}
\label{fig:Thermodynamics}
\end{figure}

In figure~\ref{fig:Thermodynamics}a, we plot $\frac{\Delta R}{R}$ as a function of fluence F at a pump-probe delay of 0.2 ps and at 1.74 eV, a representative photon-energy. As mentioned above, this quantity scales linearly with the fluence below F$_2$ (5.1 mJ cm$^{-2}$). Above F$_2$ it saturates, in correspondence with the switching-like behavior (see figure~\ref{fig:dRR}c). Second, we plot $\frac{\Delta R}{R}$ at 8 ps, again at 1.74 eV as a function of fluence (see figure~\ref{fig:Thermodynamics}b, blue curve).
In this case there are two characteristic fluences. Below F$_1$ the system quickly relaxes to a thermal state, as shown in figure~\ref{fig:dRR}a, and $\frac{\Delta R}{R}(\text{1.74 eV})$ at 8 ps follows the same linear behavior as it does at 0.2 ps. Between F$_1$ and F$_2$ the long-time response departs from the linear scaling of the low fluence regime, i.e., it displays the non-linear slow behavior (see figure~\ref{fig:dRR}b). Finally, above F$_2$ the saturation of $\frac{\Delta R}{R}$, occurring when the switching-like behavior is reached, is also present in this plot (figure~\ref{fig:dRR}b).

These fluences correspond to characteristic temperatures relevant in equilibrium thermodynamic and optical data.
In fact, assuming a thermal state of the system, we can calculate its effective temperature after the delivery of energy by one pump pulse (for details of the calculation,
see the supplemental materials~\cite{Supplemental}). A pump pulse with F$_1$ delivers an amount of energy which heats the system from its base temperature of 35 K to T$_V^-$, i.e., right to the Verwey transition temperature without supplying any latent heat. Above F$_1$, the surplus of energy triggers the phase transition and the dynamic nucleation of the metallic phase, as inferred from the x-ray diffraction data by de Jong et al.~\cite{deJong2013}. The system is brought to T$_V^+$, i.e., supplied with the full latent heat, by F$_1^+$= 4.1 mJ cm$^{-2}$. We will discuss the role of this fluence in the next section, showing how the qualitative change of behavior which becomes manifest above F$_2$ already starts for F $>$ F$_1^+$. A pump fluence of F$_2$ heats the system to an effective temperature of about 140 K. At this temperature the fluctuations towards charge ordering are still clearly visible in the equilibrium optical properties (see the inset of the figure~\ref{fig:OpticalEquilibrium}).
We argue that above F$_2$ the electronic properties of the sample undergo a sudden change and the nucleation process is no longer obervable in the out-of-equilibrium optical properties. In fact, above F$_2$ $\frac{\Delta R}{R}$ progressively loses all the characteristic features of the lower fluences. We will further support this hypothesis in the following sections.

In the inset of figure~\ref{fig:Thermodynamics}a, we compare the calculated fluences corresponding to the mentioned final temperatures for initial temperatures between 15 K and T$_V$ with the characteristic fluences measured at 35 and 80 K. As shown, the characteristic fluences at 80 K are lower than the ones at 35 K and the outlined correspondence is fully valid also at that temperature (see the supplemental materials~\cite{Supplemental} for the details of the comparison).

\subsection{Nucleation of the metallic phase, out-of-equilibrium phase separation, and non-separability of the response}
\label{subsec:NonSeparability}

In this section we will show that the response of magnetite is homogeneous over the illuminated sample for base temperatures below T$_V$ in the low and high fluence regimes, while the dynamics triggered in the intermediate regime involves phase separation.

We will show that in the intermediate fluence regime $\frac{\Delta R}{R}(h\nu, t)$ can, and must, be written as the sum of two spectro-temporal features, i.e., two different spectral features evolving in time in two different ways. These are the signatures of the production and relaxation of excitations in the charge-ordered phase, and of the nucleation of the high-temperature phase. First of all we will identify the two components on a physical basis. Then we will use singular value decomposition to show that $\frac{\Delta R}{R}(h\nu, t)$ cannot be separated as the product of a single spectral feature and a temporal evolution, but must be written as the sum of two spectral features with two different evolutions. This will prove that the decomposition into two contributions is not an artifact of the ``physical'' analysis. Moreover, it will allow us to define a general condition for the identification of phase separation in out-of-equilibrium systems. In the next section we will analyze the two components from the spectral point of view and we will show that one of the two can indeed be associated to the nucleation of the charge-disordered phase.

\begin{figure}[t]
\centering
\includegraphics[width=0.98\linewidth]{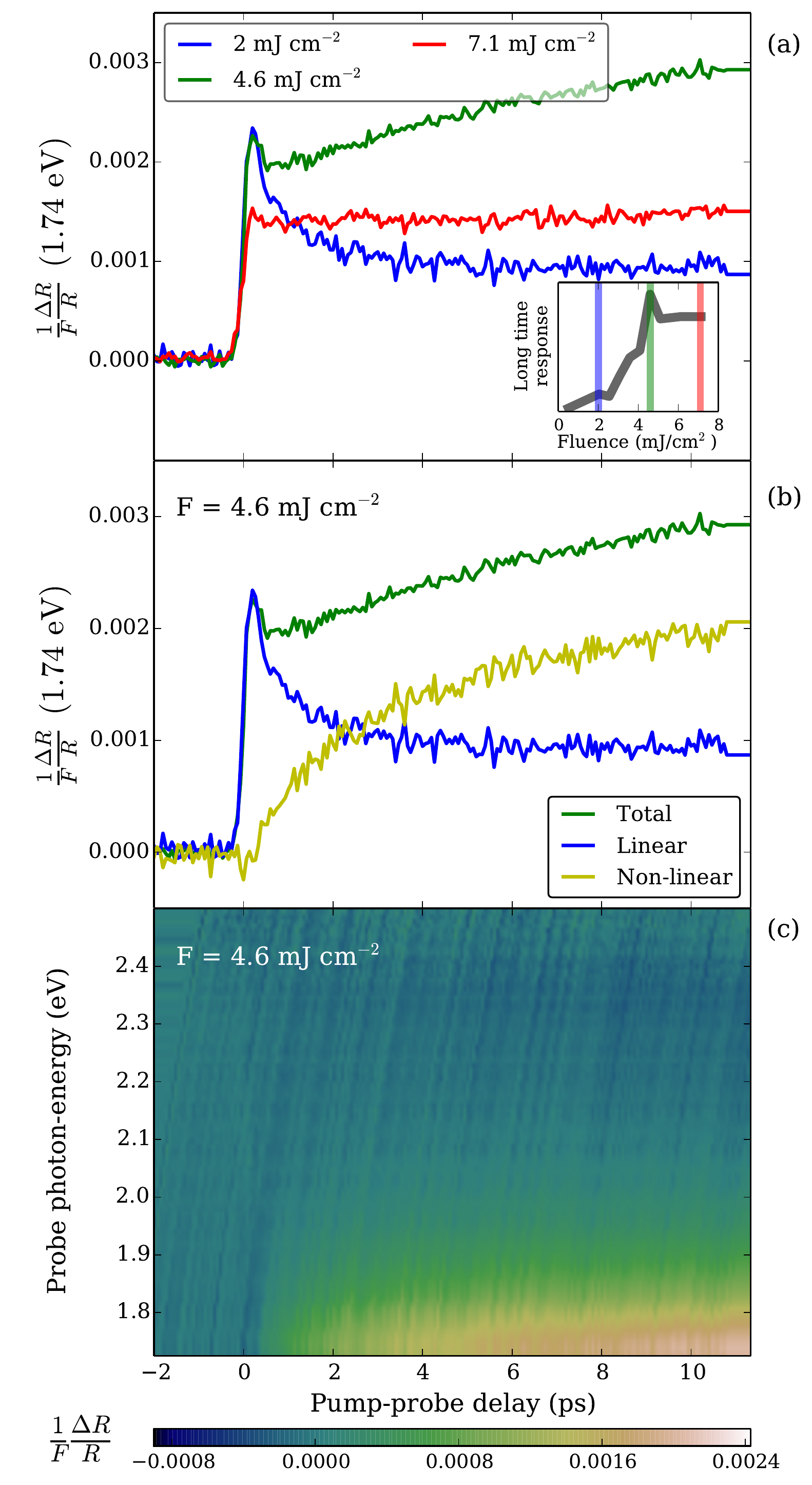}
\caption{
(a) $\frac{1}{F}\frac{\Delta R}{R}(t)$ at 1.74 eV for 2.0 mJ cm$^{-2}$ (blue), 4.6 mJ cm$^{-2}$ (green), 7.1 mJ cm$^{-2}$ (red) pump fluences. Inset: visual reference to figure~\ref{fig:Thermodynamics}b.
(b) Green: as in (a). Blue (yellow): linear (non-linear) term of $\frac{1}{F}\frac{\Delta R}{R}$ for F = 4.6 mJ cm$^{-2}$ at 1.74 eV photon-energy. 
(c) $\frac{1}{F}\frac{\Delta R}{R}'(h\nu,t)$ for F = 4.6 mJ cm$^{-2}$.
}
\label{fig:IntermediateDynamics}
\end{figure}

As a visual reference for what follows, we plot $\frac{1}{F}\frac{\Delta R}{R}$ at 1.74 eV in figure~\ref{fig:IntermediateDynamics}a for representative fluences of the three regimes. As already mentioned, $\frac{\Delta R}{R}$ scales linearly with the fluence below F$_1$. We can therefore isolate the non-linear term ($\frac{\Delta R}{R}'$) above F$_1$ by calculating
\begin{equation}\label{eq:Subtraction}
\frac{1}{F}\frac{\Delta R}{R}' = \frac{1}{F} \frac{\Delta R}{R} - \frac{1}{F_0} \frac{\Delta R}{R}\biggl|_0 \, ,
\end{equation}
where F$_0$ is the lowest fluence that we have used experimentally. In the intermediate fluence regime, this gives the result shown in figures~\ref{fig:IntermediateDynamics}b and~\ref{fig:IntermediateDynamics}c for F = 4.6 mJ cm$^{-2}$. There are three important properties of this result. First, $\frac{\Delta R}{R}'$ at t = 0.2 ps is 0 for all probe-photon energies. The non-linear $\frac{\Delta R}{R}'$ is therefore characterized by a purely slow dynamics. Second, the timescales $\tau_2$ of the non-linear/slow response match the ones obtained by de Jong et al.~\cite{deJong2013} for the emergence of regions with the high-temperature symmetry in the same excitation conditions (see supplemental material~\cite{Supplemental}). This demonstrates that the decomposition is physically meaningful, which is remarkable from a spectroscopic point of view since it entails the third important property: the total reflectivity is given by the \textit{sum} of two terms which by themselves have physical meaning. This is possible only if the total reflectivity of the sample is given by the sum of the reflectivities of distinct regions sufficiently defined to have their own optical properties. In the case of out-of-equilibrium magnetite, these are the charge-ordered insulating phase and the nucleating metallic regions. This analysis is limited to fluences between F$_1$ and F$_2$ (and trivially below F$_1$). Above F$_2$, $\frac{\Delta R}{R}$ shows a resolution-limited switch-like behavior and the subtraction described in Eq.\ \ref{eq:Subtraction} is not applicable anymore.

In order to check that this result is not just a consequence of a biased physical picture, we performed the singular value decomposition on the matrices $[\frac{\Delta R}{R}]_{h\nu,t}$, where the row and column indices are the probe-photon-energy and the pump-probe delay, respectively. The singular value decomposition algorithm is derived imposing constraints on the general problem of the factorization of a matrix, in our case representing the spectro-temporal response of the material to the excitation. These constraints may lead to singular spectro-temporal features (left and right singular vectors) which are not suitable for a physical interpretation.\footnote{In more detail, the left and right singular vectors are, separately, an orthonormal basis} However, if the number of relevant\footnote{Relevant with respect to noise.} singular values is larger than 1, the matrix is not separable in a single spectral feature evolving in time, independently of the constraints. In our data, the relevant singular values are at most two (see Supplemental material~\cite{Supplemental}, figure S3). In figure~\ref{fig:Thermodynamics}b we plot the ratio of the second largest singular value and the largest one, which gives an indication of how approximately separable $[\frac{\Delta R}{R}]_{h\nu,t}$ is and which we will refer to as ``non-separability''. When the ratio is large, $\frac{\Delta R}{R}$ is not separable, i.e., it cannot be expressed as a single spectro-temporal feature. It turns out that $\frac{\Delta R}{R}$ is the furthest from being separable in the intermediate region. Starting from zero, as the fluence increases $\frac{\Delta R}{R}$ ceases to be separable and the relevance of the additional spectro-temporal feature increases, until all the latent heat is delivered by the pump pulse. Up to this point, when the phase transition can be triggered, the pump does not deliver the energy to bring the whole illuminated sample to the high-temperature phase and the nucleation of holes in the trimeron lattice is needed to form smaller volumes fully in the high-temperature phase. Beyond that fluence (F$_1^+$ = 4.1 mJ cm$^{-2}$ as defined in the previous section), the ``non-separability'' begins to decrease and the out-of-equilibrium system becomes approximately homogeneous again for F $>$ F$_2$. In the high fluence regime, in fact, $\frac{\Delta R}{R}$ can be expressed almost as a single spectro-temporal feature.

To benchmark this procedure, we performed the singular value decomposition on the data obtained at 140 K, where we do not expect phase separation to occur, i.e., where we expect the sample to be homogeneous. The results show that $\frac{\Delta R}{R}|_{\text{140 K}}$ is exactly separable as a single spectral feature evolving in time for all the pump fluences explored. Apart from supporting the outlined picture, this leads to the consideration that a necessary condition for well-defined out-of-equilibrium phase separation is that the dynamical properties of the system are non-separable.

\subsection{Equilibrium optical properties across the phase transition}\label{subsec:EquilibriumDiscussion}


We now discuss the assignment of the features observed in the equilibrium optical properties to the excitations in magnetite. For this purpose it is useful to consider the LSDA+U results of Leonov and collaborators.\cite{Leonov2006} Charge-transfer excitations from O$_{2p}$ states to Fe$_{3d}$ states typically show much more spectral weight than transitions between Fe$_{3d}$ states. These strong charge-transfer excitations set in at about 2.5\,eV.\cite{Leonov2006,Park1998} Accordingly, the features at 0.6\,eV and 2\,eV can be attributed to excitations between Fe$_{3d}$ states. Since there are A and B sites as well as Fe$^{2+}$ $(3d^6)$ and Fe$^{3+}$ $(3d^5)$ ions, there is a multitude of possible excitations. However, the $3d^5$ configuration with only parallel spins ($S$\,=\,5/2) is very stable, hence intersite excitations of the type $3d_i^5\, 3d_j^5 \to 3d_i^4 \, 3d_j^6$ in which an electron is hopping from site $i$ to site $j$ are typically observed above 3\,eV.\cite{Reul2013,Pisarev2009} Therefore, it is reasonable to assume that the two features at 0.6\,eV and 2\,eV correspond to $3d_i^6\, 3d_j^5 \to 3d_i^5 \, 3d_j^6$ excitations.

Among the occupied states, the minority $\downarrow$ electron within the $t_{2g}$ level of an Fe$_B^{2+}$ site is closest to the Fermi level $E_F$, while the lowest unoccupied states were identified as the empty $t_{2g}$ $\downarrow$ states of Fe$_B^{3+}$ sites.\cite{Leonov2006} The peak at 0.6\,eV can thus be attributed to an intersite excitation involving these two states.\cite{Park1998,Leonov2006,Gasparov2000} This excitation gains spectral weight with increasing charge disproportionation between neighbouring Fe$_B$ sites, i.e., upon cooling towards the Verwey transition to the charge-ordered state.

In Fe$_3$O$_4$, neighbouring Fe$_B$ sites are connected via $90^\circ$ Fe-O-Fe bonds, which allow not only for intersite $t_{2g} \to t_{2g}$ hopping relevant for the excitation at 0.6\,eV but also for intersite $t_{2g} \to e_{g}$ hopping (see, e.g., Ref.\ [\onlinecite{Khomskii2014}]). According to LSDA+U results,\cite{Leonov2006} the intersite excitation of the minority $\downarrow$ $t_{2g}$ electron from an Fe$_B^{2+}$ site to the empty $e_g$ $\downarrow$ states on an Fe$_B^{3+}$ site is located at about 2\,eV.\@ The temperature dependence of the spectral weight of this excitation is more subtle. At first sight, one may expect the same temperature dependence as for the 0.6\,eV feature, since in both excitations the minority $\downarrow$ electron from an Fe$_B^{2+}$ site is hopping to a $\downarrow$ state on an Fe$_B^{3+}$ site. This disagrees with our experimental result which shows the opposite temperature dependence for the two features at 0.6\,eV and 2\,eV, see figure~\ref{fig:SW}. However, Leonov \textit{et al.}\cite{Leonov2006} find that charge order is strongly screened by a change of covalency, i.e., below $T_V$ Fe$_B^{3+}$ sites show an \textit{enhanced} occupation of the $e_g$ $\downarrow$ levels arising from hybridization with O$_{2p}$ states. Accordingly, the spectral weight for excitations into the $e_g$ levels is \textit{reduced} upon entering the charge-ordered state.

Due to the Pauli principle, the minority $\downarrow$ electron from an Fe$_B$ site may not hop to an Fe$_A$ site where all $\downarrow$ states within the $3d$ shell are occupied. Alternatively, it has been proposed\cite{Park1998,Kim2007} that the peak at 2\,eV corresponds to excitations of a majority $\uparrow$ electron from a Fe$_B^{2+}$ $(3d^6)$ site to an empty $e_g$ level at an Fe$_A^{3+}$ site. According to LSDA+U results, this excitation is expected at a slightly larger energy than the one described above.\cite{Leonov2006}

\subsection{Spectral response of the phase transition}
On the basis of the above discussion we bring our final evidence to support the proposed discussion of the photo-induced phase transition. We present the spectral analysis of the features appearing in $\frac{\Delta R}{R}$. It will allow us to show that they are indeed associated to the charge-ordered and the charge-disordered phases. 
To do this we will resort  to fits of $\frac{\Delta R}{R}$ by changing parameters of the oscillator model of the equilibrium optical properties. From such fits we can obtain two kinds of information. The first is the minimal set of free parameters (or oscillators) needed to account for the observed variation of the reflectivity. The second kind of information is the temporal evolution of the free parameters. We will restrict our discussion to the evolution of the spectral weights of the oscillators which is the most reliable outcome of the fits.\\

\begin{figure}[t]
\centering
\includegraphics[width=\linewidth]{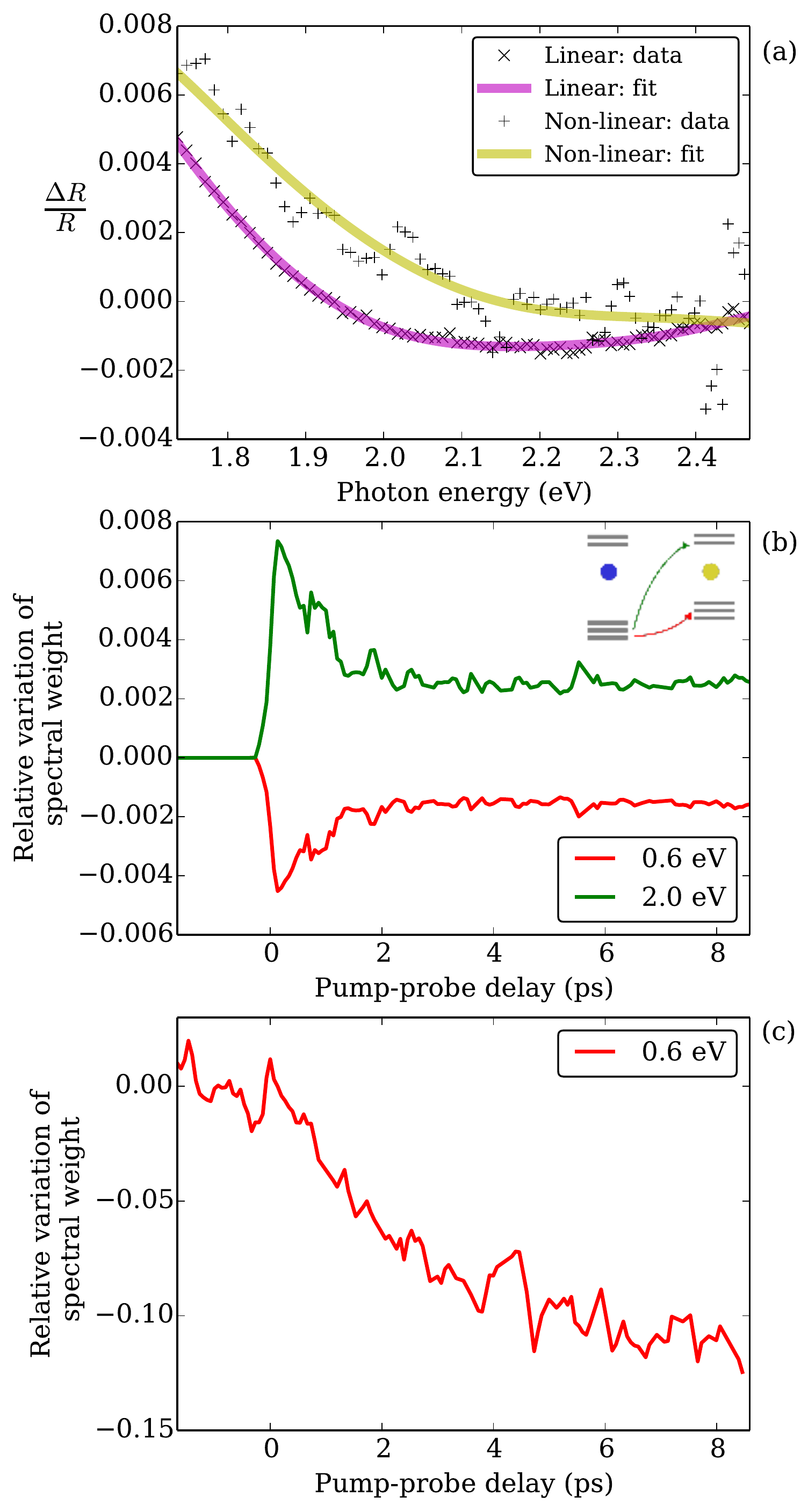}
\caption{
(a) Linear and non-linear terms in $\frac{\Delta R}{R}$ at 0.2 ps and 8.0 ps respectively, and their fits.
(b) Variation of the spectral weight as a function of pump-probe delay of the 0.6 (red) and 2.0 (green) oscillators as resulting from the fit of the linear/low-fluence $\frac{\Delta R}{R}$. Inset: sketch of the involved transitions.
(c) Variation of the spectral weight of the 0.6 eV oscillator from the fit of the non-linear term of $\frac{\Delta R}{R}$.
}
\label{fig:Spectral}
\end{figure}

In the low fluence regime (F $<$ F$_1$), the $\frac{\Delta R}{R}$ in the measured spectral range can be fully described by a variation of the oscillator in the infrared (0.6 eV), arising from Fe$_B^{2+}$t$_{\text{2g}}\rightarrow$ Fe$_B^{3+}$t$_{\text{2g}}$ transitions, and the oscillator centred at 2 eV corresponding to the Fe$_B^{2+}$t$_{\text{2g}}\rightarrow$ Fe$_B^{3+}$e$_{\text{g}}$ transitions.~\cite{Leonov2006} The fit is shown in figure~\ref{fig:Spectral}a for 0.2 ps pump-probe delay. Modifications of the charge-transfer excitations between O$_{\text{2p}}$ and Fe$_{\text{3d}}$ are, instead, not needed to describe the observed dynamical response. In figure~\ref{fig:Spectral}b we plot the temporal evolution of the spectral weights of the involved oscillators (for more details about the fitting procedure, see the supplemental material~\cite{Supplemental}). As excitations are created in the system at t = 0, the spectral weight of the Fe$_B^{2+}$t$_{\text{2g}}\rightarrow$ Fe$_B^{3+}$t$_{\text{2g}}$ (0.6 eV) transition decreases while the one of the Fe$_B^{2+}$t$_{\text{2g}}\rightarrow$ Fe$_B^{3+}$e$_{\text{g}}$ transition (2 eV) increases, then relaxing to a thermal plateau. The opposite signs of these variations are consistent with the opposite temperature behavior of the two features observed in the equilibrium data, see figure~\ref{fig:SW}. On one side the partial destruction of the charge order reduces the spectral weight of the 0.6 eV oscillator, while on the other the de-hybridization of the Fe$_B^{3+}$e$_{\text{g}}$ and O$_{\text{2p}}$ states increases the weight of the 2 eV oscillator.

We then analyze the non-linear term $\frac{\Delta R}{R}'$ arising in the intermediate fluence regime, linked to the nucleating phase. As shown in figure~\ref{fig:Spectral}a, it can be accounted for simply by a change of the Fe$_B^{2+}$t$_{\text{2g}}\rightarrow$ Fe$_B^{2+}$t$_{\text{2g}}$ transition (0.6 eV oscillator). Its spectral weight decreases with the timescale characteristic of the slow dynamics (figure~\ref{fig:Spectral}c), consistently with the picture in which the nucleating phase is the charge-disordered one.

These results suggest that the hybridization of Fe$_B^{3+}$e$_{\text{g}}$ and O$_{\text{2p}}$ states is not involved in the nucleation process, but is exclusively linked to the increase of the temperature of the system.

\section{Conclusions}
We reported measurements of both equilibrium and out-of-equilibrium optical properties of magnetite on a broad spectral range and at different temperatures across the Verwey insulator-to-metal phase transition. 
The equilibrium optical properties show a step-like behavior at the transition between the charge-ordered and charge-disordered phases. Our measurements allowed us also to determine the behavior of the spectroscopic features as a function of temperature. The most important ones in this discussion are the intersite transitions of minority spins $\downarrow$ from the Fe$_B^{2+}$t$_{\text{2g}}$ levels to the t$_{\text{2g}}$ and e$_{\text{g}}$ levels of Fe$_B^{3+}$ atoms. As expected, the spectral weight of the Fe$_B^{2+}$t$_{\text{2g}}\rightarrow$Fe$_B^{3+}$t$_{\text{2g}}$ oscillator grows upon cooling, i.e. upon increasing charge disproportionation. The temperature dependence of the spectral weight of the Fe$_B^{2+}$t$_{\text{2g}}\rightarrow$Fe$_B^{3+}$e$_{\text{g}}$ oscillator is instead opposite. It decreases upon cooling as charge ordering enhances the hybridization of Fe$_B^{3+}$ e$_g$ with O$_{2p}$ orbitals and hence gives rise to  an increased minority spin $\downarrow$ occupation of the Fe$_B^{3+}$e$_{\text{g}}$ states.

The out-of-equilibrium data allowed us to draw various conclusions on the observed dynamics. Its excitation fluence dependence reveals that the photo-excitation process can trigger the out-of-equilibrium transition analogous of the Verwey phase transition, as already reported by de Jong et al.~\cite{deJong2013}. Below a certain threshold fluence (F $<$ F$_1$, \textit{low fluence} regime), the dynamical response we observe is the one associated to a warmer charge-ordered lattice, homogeneous over the sample. With larger fluences (F$_1$ $<$ F $<$ F$_2$, \textit{intermediate fluence} regime), the high temperature phase can nucleate, eventually leading to isolated remnants of the charge-ordered lattice.~\cite{deJong2013} This picture of the nature of the nucleating phase is supported by the spectral analysis of our out-of-equilibrium data. Moreover, the latter contain also indications about the phase separation occurring in the sample. In fact, in the intermediate fluence regime $[\frac{\Delta R}{R}]_{h\nu,t}$ is not a separable matrix, i.e. it cannot be expressed as a single spectral feature evolving in time. This points to the fact that the observed response is the sum of the responses of distinct regions. Finally, we have shown that above a further threshold fluence (F $>$ F$_2$, \textit{high fluence} regime), the transition to the high temperature phase is homogeneous over the sample and nucleation is not observed in the electronic properties as it is in the intermediate fluence regime. All the mentioned characteristic fluences are linked to the equilibrium thermodynamics of magnetite and in particular to the delivery of latent heat to the sample.

Our results about the phase separation in the system may have a general relevance beyond the particular case-study of magnetite. Although the details as the lattice order and the timescales involved could be different, the behavior we discussed in this work may be valid in general for photo-excited out-of-equilibrium systems displaying a first order phase transition. Furthermore, the picture emerging from this work suggests also that a non-separable dynamical response may be a general fingerprint of out-of-equilibrium phase separation and may represent a straightforward way to identify phase separation in other out-of-equilibrium experiments.

\section*{Acknowledgements}
F. Randi and I. Vergara equally contributed to the present work.
The authors are grateful to Daniel Khomskii for the insightful discussion and critical reading of the manuscript, and to Christian Sch\"{u}{\ss}ler-Langeheine for the discussion and the sample characterization.
Research at the Universit\`{a} degli Studi di Trieste was supported through the Friuli-Venezia Giulia region (European social fund, operative program 2007/2013).
Research at Stanford was supported through the Stanford Institute for Materials and Energy Sciences (SIMES) under contract DE-AC02-76SF00515  by the US Department of Energy, Office of Basic Energy Sciences.

\bibliography{magnetite}

\begin{thebibliography}{36}%
\makeatletter
\providecommand \@ifxundefined [1]{%
 \@ifx{#1\undefined}
}%
\providecommand \@ifnum [1]{%
 \ifnum #1\expandafter \@firstoftwo
 \else \expandafter \@secondoftwo
 \fi
}%
\providecommand \@ifx [1]{%
 \ifx #1\expandafter \@firstoftwo
 \else \expandafter \@secondoftwo
 \fi
}%
\providecommand \natexlab [1]{#1}%
\providecommand \enquote  [1]{``#1''}%
\providecommand \bibnamefont  [1]{#1}%
\providecommand \bibfnamefont [1]{#1}%
\providecommand \citenamefont [1]{#1}%
\providecommand \href@noop [0]{\@secondoftwo}%
\providecommand \href [0]{\begingroup \@sanitize@url \@href}%
\providecommand \@href[1]{\@@startlink{#1}\@@href}%
\providecommand \@@href[1]{\endgroup#1\@@endlink}%
\providecommand \@sanitize@url [0]{\catcode `\\12\catcode `\$12\catcode
  `\&12\catcode `\#12\catcode `\^12\catcode `\_12\catcode `\%12\relax}%
\providecommand \@@startlink[1]{}%
\providecommand \@@endlink[0]{}%
\providecommand \url  [0]{\begingroup\@sanitize@url \@url }%
\providecommand \@url [1]{\endgroup\@href {#1}{\urlprefix }}%
\providecommand \urlprefix  [0]{URL }%
\providecommand \Eprint [0]{\href }%
\providecommand \doibase [0]{http://dx.doi.org/}%
\providecommand \selectlanguage [0]{\@gobble}%
\providecommand \bibinfo  [0]{\@secondoftwo}%
\providecommand \bibfield  [0]{\@secondoftwo}%
\providecommand \translation [1]{[#1]}%
\providecommand \BibitemOpen [0]{}%
\providecommand \bibitemStop [0]{}%
\providecommand \bibitemNoStop [0]{.\EOS\space}%
\providecommand \EOS [0]{\spacefactor3000\relax}%
\providecommand \BibitemShut  [1]{\csname bibitem#1\endcsname}%
\let\auto@bib@innerbib\@empty
\bibitem [{\citenamefont {{V}erwey}(1939)}]{Verwey1939}%
  \BibitemOpen
  \bibfield  {author} {\bibinfo {author} {\bibfnamefont {E.~J.~W.}\
  \bibnamefont {{V}erwey}},\ }\href@noop {} {\bibfield  {journal} {\bibinfo
  {journal} {Nature (London)}\ }\textbf {\bibinfo {volume} {144}},\ \bibinfo
  {pages} {327} (\bibinfo {year} {1939})}\BibitemShut {NoStop}%
\bibitem [{\citenamefont {Khomskii}(2014)}]{Khomskii2014}%
  \BibitemOpen
  \bibfield  {author} {\bibinfo {author} {\bibfnamefont {D.~I.}\ \bibnamefont
  {Khomskii}},\ }\href@noop {} {\emph {\bibinfo {title} {Transition metal
  compounds}}}\ (\bibinfo  {publisher} {Cambridge},\ \bibinfo {year}
  {2014})\BibitemShut {NoStop}%
\bibitem [{\citenamefont {Senn}\ \emph {et~al.}(2012)\citenamefont {Senn},
  \citenamefont {Wright},\ and\ \citenamefont {Attfield}}]{Senn2012}%
  \BibitemOpen
  \bibfield  {author} {\bibinfo {author} {\bibfnamefont {M.~S.}\ \bibnamefont
  {Senn}}, \bibinfo {author} {\bibfnamefont {J.~P.}\ \bibnamefont {Wright}}, \
  and\ \bibinfo {author} {\bibfnamefont {J.~P.}\ \bibnamefont {Attfield}},\
  }\href@noop {} {\bibfield  {journal} {\bibinfo  {journal} {Nature (London)}\
  }\textbf {\bibinfo {volume} {481}},\ \bibinfo {pages} {173} (\bibinfo {year}
  {2012})}\BibitemShut {NoStop}%
\bibitem [{\citenamefont {Nazarenko}\ \emph {et~al.}(2006)\citenamefont
  {Nazarenko}, \citenamefont {Lorenzo}, \citenamefont {Joly}, \citenamefont
  {Hodeau}, \citenamefont {Mannix},\ and\ \citenamefont
  {Marin}}]{Nazarenko2006}%
  \BibitemOpen
  \bibfield  {author} {\bibinfo {author} {\bibfnamefont {E.}~\bibnamefont
  {Nazarenko}}, \bibinfo {author} {\bibfnamefont {J.~E.}\ \bibnamefont
  {Lorenzo}}, \bibinfo {author} {\bibfnamefont {Y.}~\bibnamefont {Joly}},
  \bibinfo {author} {\bibfnamefont {J.~L.}\ \bibnamefont {Hodeau}}, \bibinfo
  {author} {\bibfnamefont {D.}~\bibnamefont {Mannix}}, \ and\ \bibinfo {author}
  {\bibfnamefont {C.}~\bibnamefont {Marin}},\ }\href@noop {} {\bibfield
  {journal} {\bibinfo  {journal} {Phys. Rev. Lett.}\ }\textbf {\bibinfo
  {volume} {97}},\ \bibinfo {pages} {066403} (\bibinfo {year}
  {2006})}\BibitemShut {NoStop}%
\bibitem [{\citenamefont {Lorenzo}\ \emph {et~al.}(2008)\citenamefont
  {Lorenzo}, \citenamefont {Mazzoli}, \citenamefont {Jaouen}, \citenamefont
  {Detlefs}, \citenamefont {Mannix}, \citenamefont {Grenier}, \citenamefont
  {Joly},\ and\ \citenamefont {Marin}}]{Lorenzo2008}%
  \BibitemOpen
  \bibfield  {author} {\bibinfo {author} {\bibfnamefont {J.~E.}\ \bibnamefont
  {Lorenzo}}, \bibinfo {author} {\bibfnamefont {C.}~\bibnamefont {Mazzoli}},
  \bibinfo {author} {\bibfnamefont {N.}~\bibnamefont {Jaouen}}, \bibinfo
  {author} {\bibfnamefont {C.}~\bibnamefont {Detlefs}}, \bibinfo {author}
  {\bibfnamefont {D.}~\bibnamefont {Mannix}}, \bibinfo {author} {\bibfnamefont
  {S.}~\bibnamefont {Grenier}}, \bibinfo {author} {\bibfnamefont
  {Y.}~\bibnamefont {Joly}}, \ and\ \bibinfo {author} {\bibfnamefont
  {C.}~\bibnamefont {Marin}},\ }\href {\doibase 10.1103/PhysRevLett.101.226401}
  {\bibfield  {journal} {\bibinfo  {journal} {Phys. Rev. Lett.}\ }\textbf
  {\bibinfo {volume} {101}},\ \bibinfo {pages} {226401} (\bibinfo {year}
  {2008})}\BibitemShut {NoStop}%
\bibitem [{\citenamefont {Garcia}\ \emph {et~al.}(2009)\citenamefont {Garcia},
  \citenamefont {Sub\'{i}as}, \citenamefont {Herrero-Mart\'{i}n}, \citenamefont
  {Blasco}, \citenamefont {Cuartero}, \citenamefont
  {Concepci\'{o}n~S\'{a}nchez}, \citenamefont {Mazzoli},\ and\ \citenamefont
  {Yakhou}}]{Garcia2009}%
  \BibitemOpen
  \bibfield  {author} {\bibinfo {author} {\bibfnamefont {J.}~\bibnamefont
  {Garcia}}, \bibinfo {author} {\bibfnamefont {G.}~\bibnamefont {Sub\'{i}as}},
  \bibinfo {author} {\bibfnamefont {J.}~\bibnamefont {Herrero-Mart\'{i}n}},
  \bibinfo {author} {\bibfnamefont {J.}~\bibnamefont {Blasco}}, \bibinfo
  {author} {\bibfnamefont {V.}~\bibnamefont {Cuartero}}, \bibinfo {author}
  {\bibfnamefont {M.}~\bibnamefont {Concepci\'{o}n~S\'{a}nchez}}, \bibinfo
  {author} {\bibfnamefont {C.}~\bibnamefont {Mazzoli}}, \ and\ \bibinfo
  {author} {\bibfnamefont {F.}~\bibnamefont {Yakhou}},\ }\href@noop {}
  {\bibfield  {journal} {\bibinfo  {journal} {Phys. Rev. Lett.}\ }\textbf
  {\bibinfo {volume} {102}},\ \bibinfo {pages} {176405} (\bibinfo {year}
  {2009})}\BibitemShut {NoStop}%
\bibitem [{\citenamefont {Weng}\ \emph {et~al.}(2012)\citenamefont {Weng},
  \citenamefont {Lee}, \citenamefont {Chen}, \citenamefont {Chu}, \citenamefont
  {Soo},\ and\ \citenamefont {Chang}}]{Weng2012}%
  \BibitemOpen
  \bibfield  {author} {\bibinfo {author} {\bibfnamefont {S.~C.}\ \bibnamefont
  {Weng}}, \bibinfo {author} {\bibfnamefont {Y.~R.}\ \bibnamefont {Lee}},
  \bibinfo {author} {\bibfnamefont {C.~G.}\ \bibnamefont {Chen}}, \bibinfo
  {author} {\bibfnamefont {C.~H.}\ \bibnamefont {Chu}}, \bibinfo {author}
  {\bibfnamefont {Y.~L.}\ \bibnamefont {Soo}}, \ and\ \bibinfo {author}
  {\bibfnamefont {S.~L.}\ \bibnamefont {Chang}},\ }\href@noop {} {\bibfield
  {journal} {\bibinfo  {journal} {Phys. Rev. Lett.}\ }\textbf {\bibinfo
  {volume} {108}},\ \bibinfo {pages} {146404} (\bibinfo {year}
  {2012})}\BibitemShut {NoStop}%
\bibitem [{\citenamefont {Blasco}\ \emph {et~al.}(2011)\citenamefont {Blasco},
  \citenamefont {Garcia},\ and\ \citenamefont {Subias}}]{Blasco2011}%
  \BibitemOpen
  \bibfield  {author} {\bibinfo {author} {\bibfnamefont {J.}~\bibnamefont
  {Blasco}}, \bibinfo {author} {\bibfnamefont {H.}~\bibnamefont {Garcia}}, \
  and\ \bibinfo {author} {\bibfnamefont {G.}~\bibnamefont {Subias}},\
  }\href@noop {} {\bibfield  {journal} {\bibinfo  {journal} {Phys. Rev. B}\
  }\textbf {\bibinfo {volume} {83}},\ \bibinfo {pages} {104105} (\bibinfo
  {year} {2011})}\BibitemShut {NoStop}%
\bibitem [{\citenamefont {Huang}\ \emph {et~al.}(2006)\citenamefont {Huang},
  \citenamefont {Lin}, \citenamefont {Okamoto}, \citenamefont {Chao},
  \citenamefont {Jeng}, \citenamefont {Guo}, \citenamefont {Hsu}, \citenamefont
  {Huang}, \citenamefont {Ling}, \citenamefont {Wu}, \citenamefont {Yang},\
  and\ \citenamefont {Chen}}]{Huang2006}%
  \BibitemOpen
  \bibfield  {author} {\bibinfo {author} {\bibfnamefont {D.~J.}\ \bibnamefont
  {Huang}}, \bibinfo {author} {\bibfnamefont {H.~J.}\ \bibnamefont {Lin}},
  \bibinfo {author} {\bibfnamefont {J.}~\bibnamefont {Okamoto}}, \bibinfo
  {author} {\bibfnamefont {K.~S.}\ \bibnamefont {Chao}}, \bibinfo {author}
  {\bibfnamefont {H.~T.}\ \bibnamefont {Jeng}}, \bibinfo {author}
  {\bibfnamefont {G.~Y.}\ \bibnamefont {Guo}}, \bibinfo {author} {\bibfnamefont
  {C.~H.}\ \bibnamefont {Hsu}}, \bibinfo {author} {\bibfnamefont {C.~M.}\
  \bibnamefont {Huang}}, \bibinfo {author} {\bibfnamefont {D.~C.}\ \bibnamefont
  {Ling}}, \bibinfo {author} {\bibfnamefont {W.~B.}\ \bibnamefont {Wu}},
  \bibinfo {author} {\bibfnamefont {C.~S.}\ \bibnamefont {Yang}}, \ and\
  \bibinfo {author} {\bibfnamefont {C.~T.}\ \bibnamefont {Chen}},\ }\href@noop
  {} {\bibfield  {journal} {\bibinfo  {journal} {Phys. Rev. Lett.}\ }\textbf
  {\bibinfo {volume} {96}},\ \bibinfo {pages} {096401} (\bibinfo {year}
  {2006})}\BibitemShut {NoStop}%
\bibitem [{\citenamefont {Schlappa}\ \emph {et~al.}(2008)\citenamefont
  {Schlappa}, \citenamefont {Sch\"{u}{\ss}ler-Langeheine}, \citenamefont
  {Chang}, \citenamefont {Ott}, \citenamefont {Tanaka}, \citenamefont {Hu},
  \citenamefont {Haverkort}, \citenamefont {Schierle}, \citenamefont {Weschke},
  \citenamefont {Kaindl},\ and\ \citenamefont {Tjeng}}]{Schlappa2008}%
  \BibitemOpen
  \bibfield  {author} {\bibinfo {author} {\bibfnamefont {J.}~\bibnamefont
  {Schlappa}}, \bibinfo {author} {\bibfnamefont {C.}~\bibnamefont
  {Sch\"{u}{\ss}ler-Langeheine}}, \bibinfo {author} {\bibfnamefont {C.~F.}\
  \bibnamefont {Chang}}, \bibinfo {author} {\bibfnamefont {H.}~\bibnamefont
  {Ott}}, \bibinfo {author} {\bibfnamefont {A.}~\bibnamefont {Tanaka}},
  \bibinfo {author} {\bibfnamefont {Z.}~\bibnamefont {Hu}}, \bibinfo {author}
  {\bibfnamefont {M.~W.}\ \bibnamefont {Haverkort}}, \bibinfo {author}
  {\bibfnamefont {E.}~\bibnamefont {Schierle}}, \bibinfo {author}
  {\bibfnamefont {E.}~\bibnamefont {Weschke}}, \bibinfo {author} {\bibfnamefont
  {G.}~\bibnamefont {Kaindl}}, \ and\ \bibinfo {author} {\bibfnamefont {L.~H.}\
  \bibnamefont {Tjeng}},\ }\href@noop {} {\bibfield  {journal} {\bibinfo
  {journal} {Phys. Rev. Lett.}\ }\textbf {\bibinfo {volume} {100}},\ \bibinfo
  {pages} {026406} (\bibinfo {year} {2008})}\BibitemShut {NoStop}%
\bibitem [{\citenamefont {Tanaka}\ \emph {et~al.}(2012)\citenamefont {Tanaka},
  \citenamefont {Chang}, \citenamefont {Buchholz}, \citenamefont {Trabant},
  \citenamefont {Schierle}, \citenamefont {Schlappa}, \citenamefont {Schmitz},
  \citenamefont {Ott}, \citenamefont {Metcalf}, \citenamefont {Tjeng},\ and\
  \citenamefont {Sch\"{u}{\ss}ler-Langeheine}}]{Tanaka2012}%
  \BibitemOpen
  \bibfield  {author} {\bibinfo {author} {\bibfnamefont {A.}~\bibnamefont
  {Tanaka}}, \bibinfo {author} {\bibfnamefont {C.~F.}\ \bibnamefont {Chang}},
  \bibinfo {author} {\bibfnamefont {M.}~\bibnamefont {Buchholz}}, \bibinfo
  {author} {\bibfnamefont {C.}~\bibnamefont {Trabant}}, \bibinfo {author}
  {\bibfnamefont {E.}~\bibnamefont {Schierle}}, \bibinfo {author}
  {\bibfnamefont {J.}~\bibnamefont {Schlappa}}, \bibinfo {author}
  {\bibfnamefont {D.}~\bibnamefont {Schmitz}}, \bibinfo {author} {\bibfnamefont
  {H.}~\bibnamefont {Ott}}, \bibinfo {author} {\bibfnamefont {P.}~\bibnamefont
  {Metcalf}}, \bibinfo {author} {\bibfnamefont {L.~H.}\ \bibnamefont {Tjeng}},
  \ and\ \bibinfo {author} {\bibfnamefont {C.}~\bibnamefont
  {Sch\"{u}{\ss}ler-Langeheine}},\ }\href@noop {} {\bibfield  {journal}
  {\bibinfo  {journal} {Phys. Rev. Lett.}\ }\textbf {\bibinfo {volume} {108}},\
  \bibinfo {pages} {227203} (\bibinfo {year} {2012})}\BibitemShut {NoStop}%
\bibitem [{\citenamefont {Wright}\ \emph {et~al.}(2002)\citenamefont {Wright},
  \citenamefont {Attfield},\ and\ \citenamefont {Radaelli}}]{Wright2002}%
  \BibitemOpen
  \bibfield  {author} {\bibinfo {author} {\bibfnamefont {J.~P.}\ \bibnamefont
  {Wright}}, \bibinfo {author} {\bibfnamefont {J.~P.}\ \bibnamefont
  {Attfield}}, \ and\ \bibinfo {author} {\bibfnamefont {P.}~\bibnamefont
  {Radaelli}},\ }\href@noop {} {\bibfield  {journal} {\bibinfo  {journal}
  {Phys. Rev. B}\ }\textbf {\bibinfo {volume} {66}},\ \bibinfo {pages} {214422}
  (\bibinfo {year} {2002})}\BibitemShut {NoStop}%
\bibitem [{\citenamefont {Fujii}\ \emph {et~al.}(1975)\citenamefont {Fujii},
  \citenamefont {Shirane},\ and\ \citenamefont {Yamada}}]{Fuji1975}%
  \BibitemOpen
  \bibfield  {author} {\bibinfo {author} {\bibfnamefont {Y.}~\bibnamefont
  {Fujii}}, \bibinfo {author} {\bibfnamefont {G.}~\bibnamefont {Shirane}}, \
  and\ \bibinfo {author} {\bibfnamefont {Y.}~\bibnamefont {Yamada}},\
  }\href@noop {} {\bibfield  {journal} {\bibinfo  {journal} {Phys. Rev. B}\
  }\textbf {\bibinfo {volume} {11}},\ \bibinfo {pages} {2036} (\bibinfo {year}
  {1975})}\BibitemShut {NoStop}%
\bibitem [{\citenamefont {Shapiro}\ \emph {et~al.}(1976)\citenamefont
  {Shapiro}, \citenamefont {Iizumi},\ and\ \citenamefont
  {Shirane}}]{Shapiro1976}%
  \BibitemOpen
  \bibfield  {author} {\bibinfo {author} {\bibfnamefont {S.~M.}\ \bibnamefont
  {Shapiro}}, \bibinfo {author} {\bibfnamefont {M.}~\bibnamefont {Iizumi}}, \
  and\ \bibinfo {author} {\bibfnamefont {G.}~\bibnamefont {Shirane}},\
  }\href@noop {} {\bibfield  {journal} {\bibinfo  {journal} {Phys. Rev. B}\
  }\textbf {\bibinfo {volume} {14}},\ \bibinfo {pages} {200} (\bibinfo {year}
  {1976})}\BibitemShut {NoStop}%
\bibitem [{\citenamefont {van~den Brink}\ and\ \citenamefont
  {Khomskii}(2008)}]{vandenBrink2008}%
  \BibitemOpen
  \bibfield  {author} {\bibinfo {author} {\bibfnamefont {J.}~\bibnamefont
  {van~den Brink}}\ and\ \bibinfo {author} {\bibfnamefont {D.~I.}\ \bibnamefont
  {Khomskii}},\ }\href {http://stacks.iop.org/0953-8984/20/i=43/a=434217}
  {\bibfield  {journal} {\bibinfo  {journal} {J. Phys.: Condens. Matter}\
  }\textbf {\bibinfo {volume} {20}},\ \bibinfo {pages} {434217} (\bibinfo
  {year} {2008})}\BibitemShut {NoStop}%
\bibitem [{\citenamefont {Shepherd}\ \emph {et~al.}(1985)\citenamefont
  {Shepherd}, \citenamefont {Koenitzer}, \citenamefont {Arag\'on},
  \citenamefont {Sandberg},\ and\ \citenamefont {Honig}}]{Sheperd1985}%
  \BibitemOpen
  \bibfield  {author} {\bibinfo {author} {\bibfnamefont {J.~P.}\ \bibnamefont
  {Shepherd}}, \bibinfo {author} {\bibfnamefont {J.~W.}\ \bibnamefont
  {Koenitzer}}, \bibinfo {author} {\bibfnamefont {R.}~\bibnamefont {Arag\'on}},
  \bibinfo {author} {\bibfnamefont {C.~J.}\ \bibnamefont {Sandberg}}, \ and\
  \bibinfo {author} {\bibfnamefont {J.~M.}\ \bibnamefont {Honig}},\ }\href
  {\doibase 10.1103/PhysRevB.31.1107} {\bibfield  {journal} {\bibinfo
  {journal} {Phys. Rev. B}\ }\textbf {\bibinfo {volume} {31}},\ \bibinfo
  {pages} {1107} (\bibinfo {year} {1985})}\BibitemShut {NoStop}%
\bibitem [{\citenamefont {de~Jong}\ \emph {et~al.}(2013)\citenamefont
  {de~Jong}, \citenamefont {Kukreja}, \citenamefont {Trabant}, \citenamefont
  {Pontius}, \citenamefont {CHang}, \citenamefont {Kachel}, \citenamefont
  {Beye}, \citenamefont {Sorgenfrei}, \citenamefont {Back} \emph
  {et~al.}}]{deJong2013}%
  \BibitemOpen
  \bibfield  {author} {\bibinfo {author} {\bibfnamefont {S.}~\bibnamefont
  {de~Jong}}, \bibinfo {author} {\bibfnamefont {R.}~\bibnamefont {Kukreja}},
  \bibinfo {author} {\bibfnamefont {C.}~\bibnamefont {Trabant}}, \bibinfo
  {author} {\bibfnamefont {N.}~\bibnamefont {Pontius}}, \bibinfo {author}
  {\bibfnamefont {C.~F.}\ \bibnamefont {CHang}}, \bibinfo {author}
  {\bibfnamefont {T.}~\bibnamefont {Kachel}}, \bibinfo {author} {\bibfnamefont
  {M.}~\bibnamefont {Beye}}, \bibinfo {author} {\bibfnamefont {F.}~\bibnamefont
  {Sorgenfrei}}, \bibinfo {author} {\bibfnamefont {B.}~\bibnamefont {Back},
  \bibfnamefont {C.~H.~Br\"{a}uer}},  \emph {et~al.},\ }\href@noop {}
  {\bibfield  {journal} {\bibinfo  {journal} {Nat. Mater.}\ }\textbf {\bibinfo
  {volume} {12}},\ \bibinfo {pages} {882} (\bibinfo {year} {2013})}\BibitemShut
  {NoStop}%
\bibitem [{\citenamefont {Cavalleri}\ \emph {et~al.}(2006)\citenamefont
  {Cavalleri}, \citenamefont {Rini},\ and\ \citenamefont
  {Schoenlein}}]{Cavalleri2006}%
  \BibitemOpen
  \bibfield  {author} {\bibinfo {author} {\bibfnamefont {A.}~\bibnamefont
  {Cavalleri}}, \bibinfo {author} {\bibfnamefont {M.}~\bibnamefont {Rini}}, \
  and\ \bibinfo {author} {\bibfnamefont {R.~W.}\ \bibnamefont {Schoenlein}},\
  }\href {\doibase 10.1143/JPSJ.75.011004} {\bibfield  {journal} {\bibinfo
  {journal} {Journal of the Physical Society of Japan}\ }\textbf {\bibinfo
  {volume} {75}},\ \bibinfo {pages} {011004} (\bibinfo {year} {2006})},\
  \Eprint {http://arxiv.org/abs/http://dx.doi.org/10.1143/JPSJ.75.011004}
  {http://dx.doi.org/10.1143/JPSJ.75.011004} \BibitemShut {NoStop}%
\bibitem [{\citenamefont {Chollet}\ \emph {et~al.}(2005)\citenamefont
  {Chollet}, \citenamefont {Guerin}, \citenamefont {Uchida}, \citenamefont
  {Fukaya}, \citenamefont {Shimoda}, \citenamefont {Ishikawa}, \citenamefont
  {Matsuda}, \citenamefont {Hasegawa}, \citenamefont {Ota}, \citenamefont
  {Yamochi}, \citenamefont {Saito}, \citenamefont {Tazaki}, \citenamefont
  {Adachi},\ and\ \citenamefont {Koshihara}}]{Chollet2005}%
  \BibitemOpen
  \bibfield  {author} {\bibinfo {author} {\bibfnamefont {M.}~\bibnamefont
  {Chollet}}, \bibinfo {author} {\bibfnamefont {L.}~\bibnamefont {Guerin}},
  \bibinfo {author} {\bibfnamefont {N.}~\bibnamefont {Uchida}}, \bibinfo
  {author} {\bibfnamefont {S.}~\bibnamefont {Fukaya}}, \bibinfo {author}
  {\bibfnamefont {H.}~\bibnamefont {Shimoda}}, \bibinfo {author} {\bibfnamefont
  {T.}~\bibnamefont {Ishikawa}}, \bibinfo {author} {\bibfnamefont
  {K.}~\bibnamefont {Matsuda}}, \bibinfo {author} {\bibfnamefont
  {T.}~\bibnamefont {Hasegawa}}, \bibinfo {author} {\bibfnamefont
  {A.}~\bibnamefont {Ota}}, \bibinfo {author} {\bibfnamefont {H.}~\bibnamefont
  {Yamochi}}, \bibinfo {author} {\bibfnamefont {G.}~\bibnamefont {Saito}},
  \bibinfo {author} {\bibfnamefont {R.}~\bibnamefont {Tazaki}}, \bibinfo
  {author} {\bibfnamefont {S.}~\bibnamefont {Adachi}}, \ and\ \bibinfo {author}
  {\bibfnamefont {S.}~\bibnamefont {Koshihara}},\ }\href {\doibase
  10.1126/science.1105067} {\bibfield  {journal} {\bibinfo  {journal}
  {Science}\ }\textbf {\bibinfo {volume} {307}},\ \bibinfo {pages} {86}
  (\bibinfo {year} {2005})},\ \Eprint
  {http://arxiv.org/abs/http://www.sciencemag.org/content/307/5706/86.full.pdf}
  {http://www.sciencemag.org/content/307/5706/86.full.pdf} \BibitemShut
  {NoStop}%
\bibitem [{\citenamefont {Fausti}\ \emph {et~al.}(2009)\citenamefont {Fausti},
  \citenamefont {Misochko},\ and\ \citenamefont {van Loosdrecht}}]{Fausti2009}%
  \BibitemOpen
  \bibfield  {author} {\bibinfo {author} {\bibfnamefont {D.}~\bibnamefont
  {Fausti}}, \bibinfo {author} {\bibfnamefont {O.~V.}\ \bibnamefont
  {Misochko}}, \ and\ \bibinfo {author} {\bibfnamefont {P.~H.~M.}\ \bibnamefont
  {van Loosdrecht}},\ }\href {\doibase 10.1103/PhysRevB.80.161207} {\bibfield
  {journal} {\bibinfo  {journal} {Phys. Rev. B}\ }\textbf {\bibinfo {volume}
  {80}},\ \bibinfo {pages} {161207} (\bibinfo {year} {2009})}\BibitemShut
  {NoStop}%
\bibitem [{\citenamefont {Borroni}\ \emph {et~al.}(2015)\citenamefont
  {Borroni}, \citenamefont {Baldini}, \citenamefont {Mann}, \citenamefont
  {Arrel}, \citenamefont {van Mourik}, \citenamefont {Teyssier}, \citenamefont
  {Lorenzana},\ and\ \citenamefont {Carbone}}]{Baldini2015}%
  \BibitemOpen
  \bibfield  {author} {\bibinfo {author} {\bibfnamefont {S.}~\bibnamefont
  {Borroni}}, \bibinfo {author} {\bibfnamefont {E.}~\bibnamefont {Baldini}},
  \bibinfo {author} {\bibfnamefont {A.}~\bibnamefont {Mann}}, \bibinfo {author}
  {\bibfnamefont {C.}~\bibnamefont {Arrel}}, \bibinfo {author} {\bibfnamefont
  {F.}~\bibnamefont {van Mourik}}, \bibinfo {author} {\bibfnamefont
  {J.}~\bibnamefont {Teyssier}}, \bibinfo {author} {\bibfnamefont
  {J.}~\bibnamefont {Lorenzana}}, \ and\ \bibinfo {author} {\bibfnamefont
  {F.}~\bibnamefont {Carbone}},\ }\href@noop {} {\  (\bibinfo {year} {2015})},\
  \Eprint {http://arxiv.org/abs/1507.07193} {arXiv:1507.07193 [cond-mat]}
  \BibitemShut {NoStop}%
\bibitem [{\citenamefont {Schlegel}\ \emph {et~al.}(1979)\citenamefont
  {Schlegel}, \citenamefont {Alvarado},\ and\ \citenamefont
  {Wachter}}]{Schlegel1979}%
  \BibitemOpen
  \bibfield  {author} {\bibinfo {author} {\bibfnamefont {A.}~\bibnamefont
  {Schlegel}}, \bibinfo {author} {\bibfnamefont {S.~F.}\ \bibnamefont
  {Alvarado}}, \ and\ \bibinfo {author} {\bibfnamefont {P.}~\bibnamefont
  {Wachter}},\ }\href@noop {} {\bibfield  {journal} {\bibinfo  {journal} {J.
  Phys. C}\ }\textbf {\bibinfo {volume} {12}},\ \bibinfo {pages} {1157}
  (\bibinfo {year} {1979})}\BibitemShut {NoStop}%
\bibitem [{\citenamefont {Gasparov}\ \emph {et~al.}(2000)\citenamefont
  {Gasparov}, \citenamefont {Tanner}, \citenamefont {Romero}, \citenamefont
  {Berger}, \citenamefont {Margaritondo},\ and\ \citenamefont
  {Forr\`{o}}}]{Gasparov2000}%
  \BibitemOpen
  \bibfield  {author} {\bibinfo {author} {\bibfnamefont {L.~V.}\ \bibnamefont
  {Gasparov}}, \bibinfo {author} {\bibfnamefont {D.~B.}\ \bibnamefont
  {Tanner}}, \bibinfo {author} {\bibfnamefont {D.~B.}\ \bibnamefont {Romero}},
  \bibinfo {author} {\bibfnamefont {H.}~\bibnamefont {Berger}}, \bibinfo
  {author} {\bibfnamefont {G.}~\bibnamefont {Margaritondo}}, \ and\ \bibinfo
  {author} {\bibfnamefont {L.}~\bibnamefont {Forr\`{o}}},\ }\href@noop {}
  {\bibfield  {journal} {\bibinfo  {journal} {Phys. Rev. B}\ }\textbf {\bibinfo
  {volume} {62}},\ \bibinfo {pages} {7939} (\bibinfo {year}
  {2000})}\BibitemShut {NoStop}%
\bibitem [{\citenamefont {Park}\ \emph {et~al.}(1998)\citenamefont {Park},
  \citenamefont {Ishikawa},\ and\ \citenamefont {Tokura}}]{Park1998}%
  \BibitemOpen
  \bibfield  {author} {\bibinfo {author} {\bibfnamefont {S.}~\bibnamefont
  {Park}}, \bibinfo {author} {\bibfnamefont {T.}~\bibnamefont {Ishikawa}}, \
  and\ \bibinfo {author} {\bibfnamefont {Y.}~\bibnamefont {Tokura}},\
  }\href@noop {} {\bibfield  {journal} {\bibinfo  {journal} {Phys. Rev. B}\
  }\textbf {\bibinfo {volume} {58}},\ \bibinfo {pages} {3717} (\bibinfo {year}
  {1998})}\BibitemShut {NoStop}%
\bibitem [{\citenamefont {Leonov}\ \emph {et~al.}(2006)\citenamefont {Leonov},
  \citenamefont {Yaresko}, \citenamefont {Antonov},\ and\ \citenamefont
  {Anisimov}}]{Leonov2006}%
  \BibitemOpen
  \bibfield  {author} {\bibinfo {author} {\bibfnamefont {I.}~\bibnamefont
  {Leonov}}, \bibinfo {author} {\bibfnamefont {A.~N.}\ \bibnamefont {Yaresko}},
  \bibinfo {author} {\bibfnamefont {V.~N.}\ \bibnamefont {Antonov}}, \ and\
  \bibinfo {author} {\bibfnamefont {V.~I.}\ \bibnamefont {Anisimov}},\ }\href
  {\doibase 10.1103/PhysRevB.74.165117} {\bibfield  {journal} {\bibinfo
  {journal} {Phys. Rev. B}\ }\textbf {\bibinfo {volume} {74}},\ \bibinfo
  {pages} {165117} (\bibinfo {year} {2006})}\BibitemShut {NoStop}%
\bibitem [{\citenamefont {Kim}\ \emph {et~al.}(2007)\citenamefont {Kim},
  \citenamefont {Choi}, \citenamefont {Lee}, \citenamefont {Lee},\ and\
  \citenamefont {Park}}]{Kim2007}%
  \BibitemOpen
  \bibfield  {author} {\bibinfo {author} {\bibfnamefont {K.~J.}\ \bibnamefont
  {Kim}}, \bibinfo {author} {\bibfnamefont {S.}~\bibnamefont {Choi}}, \bibinfo
  {author} {\bibfnamefont {H.~J.}\ \bibnamefont {Lee}}, \bibinfo {author}
  {\bibfnamefont {J.~H.}\ \bibnamefont {Lee}}, \ and\ \bibinfo {author}
  {\bibfnamefont {J.~Y.}\ \bibnamefont {Park}},\ }\href@noop {} {\bibfield
  {journal} {\bibinfo  {journal} {Solid State Comm.}\ }\textbf {\bibinfo
  {volume} {143}},\ \bibinfo {pages} {285} (\bibinfo {year}
  {2007})}\BibitemShut {NoStop}%
\bibitem [{\citenamefont {Novelli}\ \emph {et~al.}(2012)\citenamefont
  {Novelli}, \citenamefont {Fausti}, \citenamefont {Reul}, \citenamefont
  {Cilento}, \citenamefont {van Loosdrecht}, \citenamefont {Nugroho},
  \citenamefont {Palstra}, \citenamefont {Gr\"uninger},\ and\ \citenamefont
  {Parmigiani}}]{Novelli2012}%
  \BibitemOpen
  \bibfield  {author} {\bibinfo {author} {\bibfnamefont {F.}~\bibnamefont
  {Novelli}}, \bibinfo {author} {\bibfnamefont {D.}~\bibnamefont {Fausti}},
  \bibinfo {author} {\bibfnamefont {J.}~\bibnamefont {Reul}}, \bibinfo {author}
  {\bibfnamefont {F.}~\bibnamefont {Cilento}}, \bibinfo {author} {\bibfnamefont
  {P.~H.~M.}\ \bibnamefont {van Loosdrecht}}, \bibinfo {author} {\bibfnamefont
  {A.~A.}\ \bibnamefont {Nugroho}}, \bibinfo {author} {\bibfnamefont
  {T.~T.~M.}\ \bibnamefont {Palstra}}, \bibinfo {author} {\bibfnamefont
  {M.}~\bibnamefont {Gr\"uninger}}, \ and\ \bibinfo {author} {\bibfnamefont
  {F.}~\bibnamefont {Parmigiani}},\ }\href {\doibase
  10.1103/PhysRevB.86.165135} {\bibfield  {journal} {\bibinfo  {journal} {Phys.
  Rev. B}\ }\textbf {\bibinfo {volume} {86}},\ \bibinfo {pages} {165135}
  (\bibinfo {year} {2012})}\BibitemShut {NoStop}%
\bibitem [{\citenamefont {Kuipers}\ and\ \citenamefont
  {Brabers}(1976)}]{Kuipers1976}%
  \BibitemOpen
  \bibfield  {author} {\bibinfo {author} {\bibfnamefont {A.~J.~M.}\
  \bibnamefont {Kuipers}}\ and\ \bibinfo {author} {\bibfnamefont {V.~A.~M.}\
  \bibnamefont {Brabers}},\ }\href {\doibase 10.1103/PhysRevB.14.1401}
  {\bibfield  {journal} {\bibinfo  {journal} {Phys. Rev. B}\ }\textbf {\bibinfo
  {volume} {14}},\ \bibinfo {pages} {1401} (\bibinfo {year}
  {1976})}\BibitemShut {NoStop}%
\bibitem [{\citenamefont {G\"{o}ssling}\ \emph
  {et~al.}(2008{\natexlab{a}})\citenamefont {G\"{o}ssling}, \citenamefont
  {Haverkort}, \citenamefont {Benomar}, \citenamefont {Wu}, \citenamefont
  {Senff}, \citenamefont {M\"{o}ller}, \citenamefont {Braden}, \citenamefont
  {Mydosh},\ and\ \citenamefont {Gr\"{u}ninger}}]{Goessling2008a}%
  \BibitemOpen
  \bibfield  {author} {\bibinfo {author} {\bibfnamefont {A.}~\bibnamefont
  {G\"{o}ssling}}, \bibinfo {author} {\bibfnamefont {M.~W.}\ \bibnamefont
  {Haverkort}}, \bibinfo {author} {\bibfnamefont {M.}~\bibnamefont {Benomar}},
  \bibinfo {author} {\bibfnamefont {H.}~\bibnamefont {Wu}}, \bibinfo {author}
  {\bibfnamefont {D.}~\bibnamefont {Senff}}, \bibinfo {author} {\bibfnamefont
  {T.}~\bibnamefont {M\"{o}ller}}, \bibinfo {author} {\bibfnamefont
  {M.}~\bibnamefont {Braden}}, \bibinfo {author} {\bibfnamefont {J.~A.}\
  \bibnamefont {Mydosh}}, \ and\ \bibinfo {author} {\bibfnamefont
  {M.}~\bibnamefont {Gr\"{u}ninger}},\ }\href@noop {} {\bibfield  {journal}
  {\bibinfo  {journal} {Phys. Rev. B}\ }\textbf {\bibinfo {volume} {77}},\
  \bibinfo {pages} {035109} (\bibinfo {year} {2008}{\natexlab{a}})}\BibitemShut
  {NoStop}%
\bibitem [{\citenamefont {G\"{o}ssling}\ \emph
  {et~al.}(2008{\natexlab{b}})\citenamefont {G\"{o}ssling}, \citenamefont
  {Schmitz}, \citenamefont {Roth}, \citenamefont {Haverkort}, \citenamefont
  {Lorenz}, \citenamefont {Mydosh}, \citenamefont {M\"{u}ller-Hartmann},\ and\
  \citenamefont {Gr\"{u}ninger}}]{Goessling2008b}%
  \BibitemOpen
  \bibfield  {author} {\bibinfo {author} {\bibfnamefont {A.}~\bibnamefont
  {G\"{o}ssling}}, \bibinfo {author} {\bibfnamefont {R.}~\bibnamefont
  {Schmitz}}, \bibinfo {author} {\bibfnamefont {H.}~\bibnamefont {Roth}},
  \bibinfo {author} {\bibfnamefont {M.~W.}\ \bibnamefont {Haverkort}}, \bibinfo
  {author} {\bibfnamefont {T.}~\bibnamefont {Lorenz}}, \bibinfo {author}
  {\bibfnamefont {J.~A.}\ \bibnamefont {Mydosh}}, \bibinfo {author}
  {\bibfnamefont {E.}~\bibnamefont {M\"{u}ller-Hartmann}}, \ and\ \bibinfo
  {author} {\bibfnamefont {M.}~\bibnamefont {Gr\"{u}ninger}},\ }\href@noop {}
  {\bibfield  {journal} {\bibinfo  {journal} {Phys. Rev. B}\ }\textbf {\bibinfo
  {volume} {78}},\ \bibinfo {pages} {075122} (\bibinfo {year}
  {2008}{\natexlab{b}})}\BibitemShut {NoStop}%
\bibitem [{\citenamefont {Reul}\ \emph {et~al.}(2012)\citenamefont {Reul},
  \citenamefont {Nugroho}, \citenamefont {Palstra},\ and\ \citenamefont
  {Gr\"{u}ninger}}]{Reul2012}%
  \BibitemOpen
  \bibfield  {author} {\bibinfo {author} {\bibfnamefont {J.}~\bibnamefont
  {Reul}}, \bibinfo {author} {\bibfnamefont {A.~A.}\ \bibnamefont {Nugroho}},
  \bibinfo {author} {\bibfnamefont {T.~T.~M.}\ \bibnamefont {Palstra}}, \ and\
  \bibinfo {author} {\bibfnamefont {M.}~\bibnamefont {Gr\"{u}ninger}},\
  }\href@noop {} {\bibfield  {journal} {\bibinfo  {journal} {Phys. Rev. B}\
  }\textbf {\bibinfo {volume} {86}},\ \bibinfo {pages} {125128} (\bibinfo
  {year} {2012})}\BibitemShut {NoStop}%
\bibitem [{Sup()}]{Supplemental}%
  \BibitemOpen
  \href@noop {} {\enquote {\bibinfo {title} {Supplemental materials},}\
  }\BibitemShut {NoStop}%
\bibitem [{Note1()}]{Note1}%
  \BibitemOpen
  \bibinfo {note} {In more detail, the left and right singular vectors are,
  separately, an orthonormal basis}\BibitemShut {NoStop}%
\bibitem [{Note2()}]{Note2}%
  \BibitemOpen
  \bibinfo {note} {Relevant with respect to noise.}\BibitemShut {Stop}%
\bibitem [{\citenamefont {Reul}\ \emph {et~al.}(2013)\citenamefont {Reul},
  \citenamefont {Fels}, \citenamefont {Qureshi}, \citenamefont {Shportko},
  \citenamefont {Braden},\ and\ \citenamefont {Gr\"{u}ninger}}]{Reul2013}%
  \BibitemOpen
  \bibfield  {author} {\bibinfo {author} {\bibfnamefont {J.}~\bibnamefont
  {Reul}}, \bibinfo {author} {\bibfnamefont {L.}~\bibnamefont {Fels}}, \bibinfo
  {author} {\bibfnamefont {N.}~\bibnamefont {Qureshi}}, \bibinfo {author}
  {\bibfnamefont {K.}~\bibnamefont {Shportko}}, \bibinfo {author}
  {\bibfnamefont {M.}~\bibnamefont {Braden}}, \ and\ \bibinfo {author}
  {\bibfnamefont {M.}~\bibnamefont {Gr\"{u}ninger}},\ }\href@noop {} {\bibfield
   {journal} {\bibinfo  {journal} {Phys. Rev. B}\ }\textbf {\bibinfo {volume}
  {87}},\ \bibinfo {pages} {205142} (\bibinfo {year} {2013})}\BibitemShut
  {NoStop}%
\bibitem [{\citenamefont {Pisarev}\ \emph {et~al.}(2009)\citenamefont
  {Pisarev}, \citenamefont {Moskvin}, \citenamefont {Kalashnikova},\ and\
  \citenamefont {Rasing}}]{Pisarev2009}%
  \BibitemOpen
  \bibfield  {author} {\bibinfo {author} {\bibfnamefont {R.~V.}\ \bibnamefont
  {Pisarev}}, \bibinfo {author} {\bibfnamefont {A.~S.}\ \bibnamefont
  {Moskvin}}, \bibinfo {author} {\bibfnamefont {A.~M.}\ \bibnamefont
  {Kalashnikova}}, \ and\ \bibinfo {author} {\bibfnamefont {T.}~\bibnamefont
  {Rasing}},\ }\href@noop {} {\bibfield  {journal} {\bibinfo  {journal} {Phys.
  Rev. B}\ }\textbf {\bibinfo {volume} {79}},\ \bibinfo {pages} {235128}
  (\bibinfo {year} {2009})}\BibitemShut {NoStop}%
\end{thebibliography}%


\begin{thebibliography}{3}%
\makeatletter
\providecommand \@ifxundefined [1]{%
 \@ifx{#1\undefined}
}%
\providecommand \@ifnum [1]{%
 \ifnum #1\expandafter \@firstoftwo
 \else \expandafter \@secondoftwo
 \fi
}%
\providecommand \@ifx [1]{%
 \ifx #1\expandafter \@firstoftwo
 \else \expandafter \@secondoftwo
 \fi
}%
\providecommand \natexlab [1]{#1}%
\providecommand \enquote  [1]{``#1''}%
\providecommand \bibnamefont  [1]{#1}%
\providecommand \bibfnamefont [1]{#1}%
\providecommand \citenamefont [1]{#1}%
\providecommand \href@noop [0]{\@secondoftwo}%
\providecommand \href [0]{\begingroup \@sanitize@url \@href}%
\providecommand \@href[1]{\@@startlink{#1}\@@href}%
\providecommand \@@href[1]{\endgroup#1\@@endlink}%
\providecommand \@sanitize@url [0]{\catcode `\\12\catcode `\$12\catcode
  `\&12\catcode `\#12\catcode `\^12\catcode `\_12\catcode `\%12\relax}%
\providecommand \@@startlink[1]{}%
\providecommand \@@endlink[0]{}%
\providecommand \url  [0]{\begingroup\@sanitize@url \@url }%
\providecommand \@url [1]{\endgroup\@href {#1}{\urlprefix }}%
\providecommand \urlprefix  [0]{URL }%
\providecommand \Eprint [0]{\href }%
\providecommand \doibase [0]{http://dx.doi.org/}%
\providecommand \selectlanguage [0]{\@gobble}%
\providecommand \bibinfo  [0]{\@secondoftwo}%
\providecommand \bibfield  [0]{\@secondoftwo}%
\providecommand \translation [1]{[#1]}%
\providecommand \BibitemOpen [0]{}%
\providecommand \bibitemStop [0]{}%
\providecommand \bibitemNoStop [0]{.\EOS\space}%
\providecommand \EOS [0]{\spacefactor3000\relax}%
\providecommand \BibitemShut  [1]{\csname bibitem#1\endcsname}%
\let\auto@bib@innerbib\@empty
\bibitem [{\citenamefont {Park}\ \emph {et~al.}(1998)\citenamefont {Park},
  \citenamefont {Ishikawa},\ and\ \citenamefont {Tokura}}]{Park1998}%
  \BibitemOpen
  \bibfield  {author} {\bibinfo {author} {\bibfnamefont {S.}~\bibnamefont
  {Park}}, \bibinfo {author} {\bibfnamefont {T.}~\bibnamefont {Ishikawa}}, \
  and\ \bibinfo {author} {\bibfnamefont {Y.}~\bibnamefont {Tokura}},\
  }\href@noop {} {\bibfield  {journal} {\bibinfo  {journal} {Phys. Rev. B}\
  }\textbf {\bibinfo {volume} {58}},\ \bibinfo {pages} {3717} (\bibinfo {year}
  {1998})}\BibitemShut {NoStop}%
\bibitem [{\citenamefont {Gasparov}\ \emph {et~al.}(2000)\citenamefont
  {Gasparov}, \citenamefont {Tanner}, \citenamefont {Romero}, \citenamefont
  {Berger}, \citenamefont {Margaritondo},\ and\ \citenamefont
  {Forr\`{o}}}]{Gasparov2000}%
  \BibitemOpen
  \bibfield  {author} {\bibinfo {author} {\bibfnamefont {L.~V.}\ \bibnamefont
  {Gasparov}}, \bibinfo {author} {\bibfnamefont {D.~B.}\ \bibnamefont
  {Tanner}}, \bibinfo {author} {\bibfnamefont {D.~B.}\ \bibnamefont {Romero}},
  \bibinfo {author} {\bibfnamefont {H.}~\bibnamefont {Berger}}, \bibinfo
  {author} {\bibfnamefont {G.}~\bibnamefont {Margaritondo}}, \ and\ \bibinfo
  {author} {\bibfnamefont {L.}~\bibnamefont {Forr\`{o}}},\ }\href@noop {}
  {\bibfield  {journal} {\bibinfo  {journal} {Phys. Rev. B}\ }\textbf {\bibinfo
  {volume} {62}},\ \bibinfo {pages} {7939} (\bibinfo {year}
  {2000})}\BibitemShut {NoStop}%
\bibitem [{\citenamefont {Takai}\ \emph {et~al.}(1994)\citenamefont {Takai},
  \citenamefont {Akishige}, \citenamefont {Kawaji}, \citenamefont {Atake},\
  and\ \citenamefont {Sawaguchi}}]{Takai1994}%
  \BibitemOpen
  \bibfield  {author} {\bibinfo {author} {\bibfnamefont {S.}~\bibnamefont
  {Takai}}, \bibinfo {author} {\bibfnamefont {Y.}~\bibnamefont {Akishige}},
  \bibinfo {author} {\bibfnamefont {H.}~\bibnamefont {Kawaji}}, \bibinfo
  {author} {\bibfnamefont {T.}~\bibnamefont {Atake}}, \ and\ \bibinfo {author}
  {\bibfnamefont {E.}~\bibnamefont {Sawaguchi}},\ }\href@noop {} {\bibfield
  {journal} {\bibinfo  {journal} {J. Chem. Thermodyn.}\ }\textbf {\bibinfo
  {volume} {26}},\ \bibinfo {pages} {1259} (\bibinfo {year}
  {1994})}\BibitemShut {NoStop}%
\end{thebibliography}%

\end{document}


\title{Phase separation in the non-equilibrium Verwey transition in magnetite}

\author{F. Randi$^{**}$}%
\affiliation{Department of Physics, Universit\`{a} degli Studi di Trieste, 34127 Trieste, Italy}

\author{I. Vergara$^{**}$}%
\affiliation{II. Physikalisches Institut, Universit\"{a}t zu K\"{o}ln, 50937 K\"{o}ln, Germany}%

\author{F. Novelli}%
\affiliation{Sincrotrone Trieste SCpA, 34127 Basovizza, Italy}%

\author{M. Esposito}%
\affiliation{Department of Physics, Universit\`{a} degli Studi di Trieste, 34127 Trieste, Italy}

\author{M. Dell'Angela}%
\affiliation{Sincrotrone Trieste SCpA, 34127 Basovizza, Italy}%

\author{V. A. M. Brabers}%
\affiliation{Department of Physics, Eindhoven University of Technology, Eindhoven, The Netherlands}


\author{P. Metcalf}%
\affiliation{Purdue University, School of Materials Engineering, West Lafayette, Indiana 47907, USA}%

\author{R. Kukreja}%
\affiliation{Stanford Institute for Energy and Materials Sciences, SLAC National Accelerator Laboratory, 2575 Sand Hill Road, Menlo Park, California 94025, USA}%

\author{H. A. D\"{u}rr}%
\affiliation{Stanford Institute for Energy and Materials Sciences, SLAC National Accelerator Laboratory, 2575 Sand Hill Road, Menlo Park, California 94025, USA}%

\author{D. Fausti}%
\email[Corresponding author: ]{daniele.fausti@elettra.eu}
\affiliation{Department of Physics, Universit\`{a} degli Studi di Trieste, 34127 Trieste, Italy}
\affiliation{Sincrotrone Trieste SCpA, 34127 Basovizza, Italy}%

\author{M. Gr\"{u}ninger}
\affiliation{II. Physikalisches Institut, Universit\"{a}t zu K\"{o}ln, 50937 K\"{o}ln, Germany}%

\author{F. Parmigiani}%
\affiliation{Department of Physics, Universit\`{a} degli Studi di Trieste, 34127 Trieste, Italy}%
\affiliation{Sincrotrone Trieste SCpA, 34127 Basovizza, Italy}%
\affiliation{International Faculty, Universit\"{a}t zu K\"{o}ln, 50937 K\"{o}ln, Germany}

\date{\today}

\pacs{72.80.Ga, 78.20.Ci, 78.47.jg}
\maketitle

\section{Equilibrium optical properties}
\begin{figure}
\centering
\includegraphics[width=0.9\linewidth]{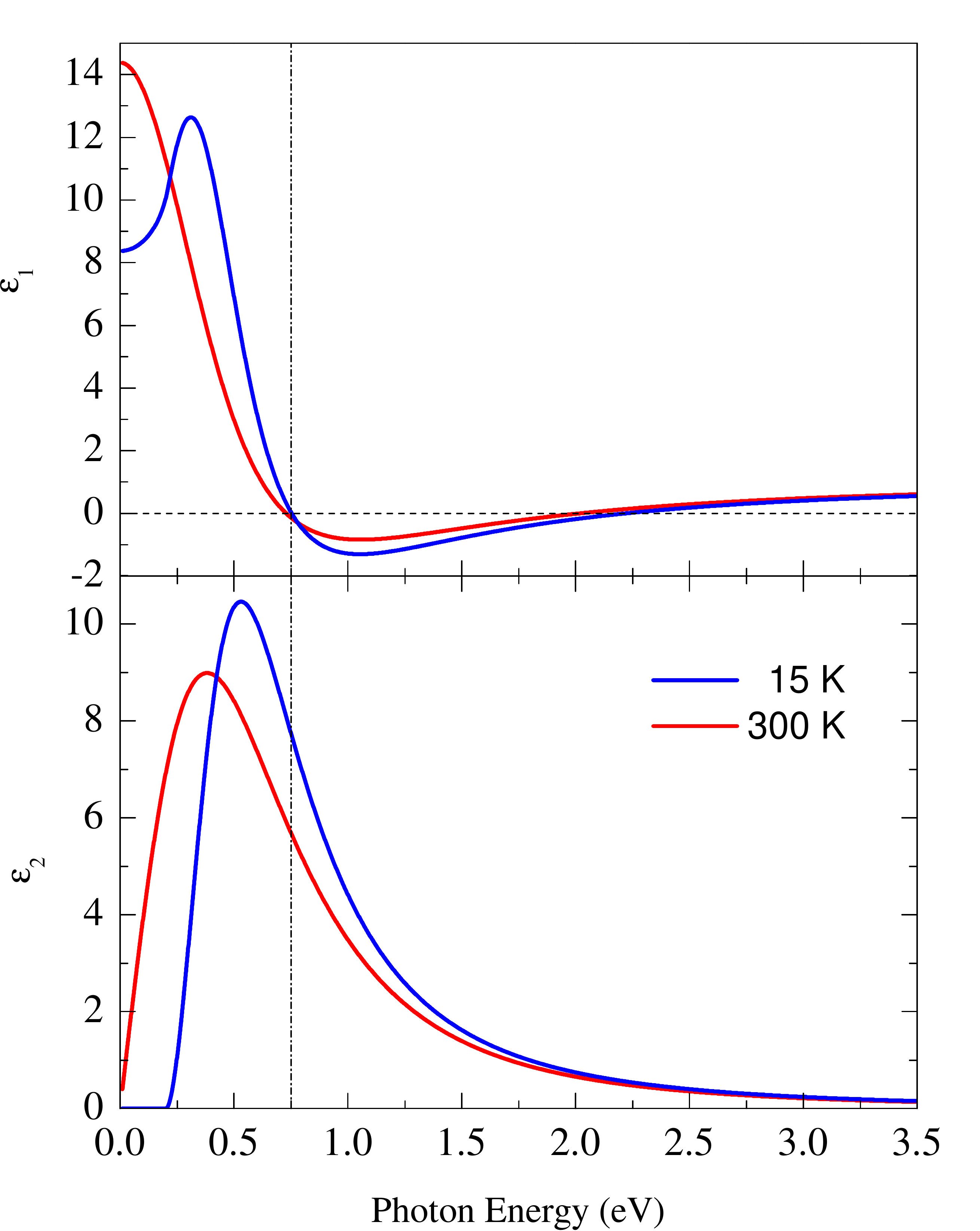}
\caption{
Real (a) and imaginary (b) parts of the dielectric function of the sole oscillator centred at 0.6 eV (Tauc-Lorentz lineshape). At 15 K the presence of a gap in $\epsilon_2$ gives rise to the formation of a peak in $\epsilon_1$ at about 0.3 eV. Vertical dashed line: lower limit of the measured photon-energy range.
}
\label{fig:SEQ1}
\end{figure}

The Tauc-Lorentz oscillator describing the peak in $\sigma_1$ at 0.6 eV mainly lies below the lower limit of our experiment, see figure~\ref{fig:SEQ1}. However, ellipsometry determines both $\epsilon_1$ and $\epsilon_2$ at each measured photon energy, thus the measured data impose a strong constraint on the lineshape of this feature in the modeling. To demonstrate that the contribution of this oscillator is distinctive in $\epsilon_1$ and $\epsilon_2$ up to high energies, we plot this feature separately in figure~\ref{fig:SEQ1}. Assuming a Tauc-Lorentz line shape, we obtain a plausible description of the experimental data in reasonable agreement with  data from the literature obtained by a Kramers-Kronig analysis of reflectivity data~\cite{Park1998,Gasparov2000}.

\section{Thermodynamics and equivalent fluence}
We studied the correspondence between the pump fluence and the energy needed to adiabatically cross the Verwey transition, i.e. the final effective temperature of the sample as a function of the pump fluence. To do this, we considered a uniformly excited volume given by the area illuminated by the pump beam times the penetration depth $\alpha$ for the pump frequency. The equivalent fluence $F(T)$ is therefore
\begin{equation}
	F(T) = e(T) \alpha \frac{1}{1-e^{-1}} \frac{1}{1-R}
\end{equation}
where $e(T)$ is the energy per unit volume needed to reach the transition (as calculated with the specific heat data measured by~\cite{Takai1994}), $1-e^{-1}$ is the fraction of energy dropped within the penetration depth, and $1-R$ the fraction of energy entering the sample.

\section{Out-of-equilibrium optical properties}
\subsection{Results at 35 K}
\begin{figure}
\centering
\includegraphics[width=0.9\linewidth]{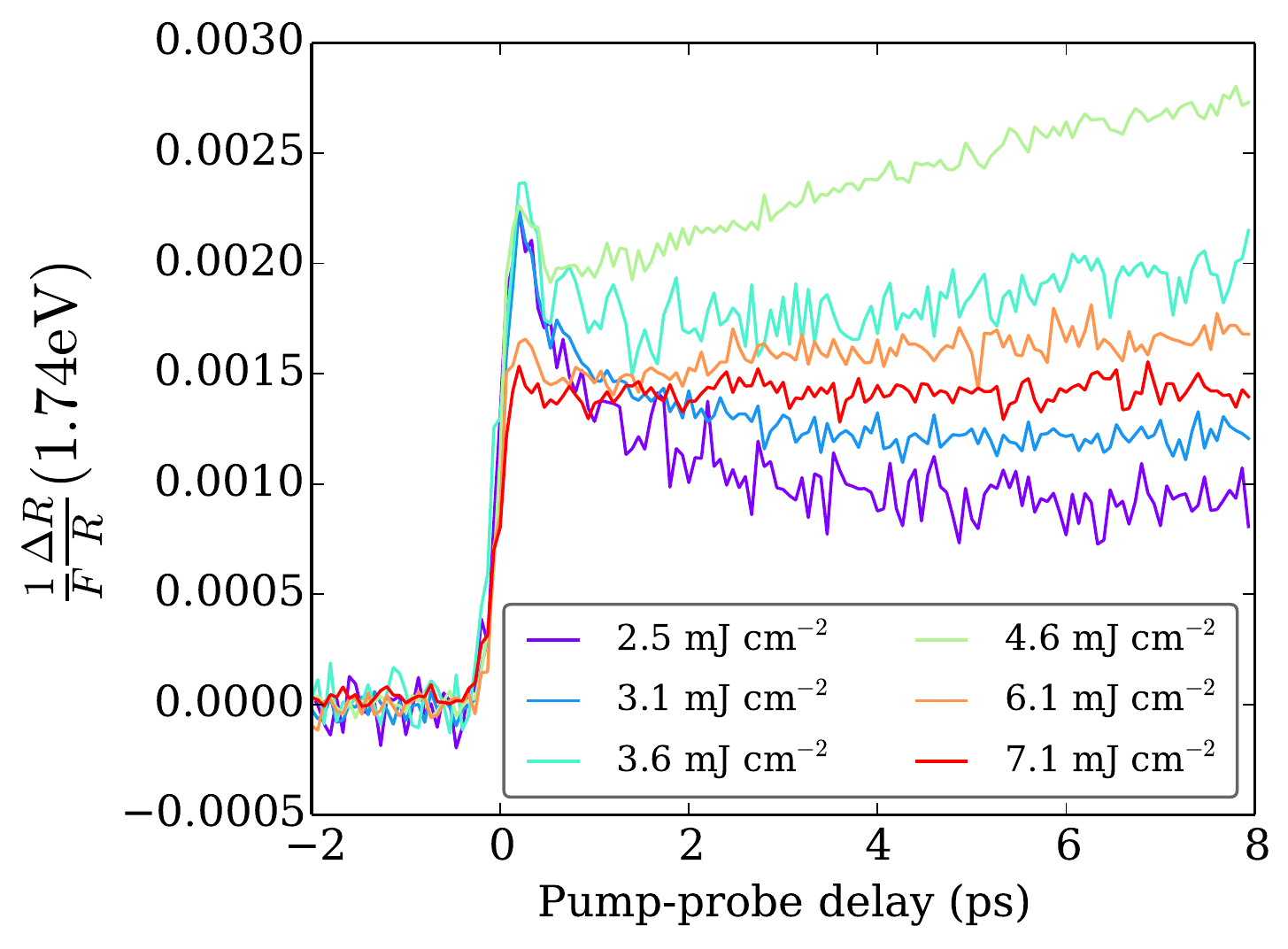}
\caption{Normalized relative variation of the reflectivity $\frac{1}{F}\frac{\Delta R}{R}$(1.74 eV) measured at 35 K for different pump fluences.
}
\label{fig:S35A}
\end{figure}

\paragraph{Fluence dependence} To show the effect of the pump fluence in more detail, in figure~\ref{fig:S35A} we plot $\frac{1}{F}\frac{\Delta R}{R}|_{\text{35 K}}(t)$ at 1.74 eV over the entire explored pump fluence range. In the low fluence regime (F $<$ F$_1$ = 2.7 mJ cm$^{-2}$) the system does not display any long-time dynamics. Above F$_1$, in the intermediate fluence regime, the nucleation of the high temperature phase is triggered and $\frac{\Delta R}{R}$ displays the slow dynamics associated to it. Finally, above F$_2$ = 5.1 mJ cm$^{-2}$ the variation of the reflectivity progressively becomes a switch. 

\paragraph{Characteristic timescales} The characteristic timescales $\tau_2$ for the slow response in the intermediate fluence regime at 35 K are 5.6 ps (F = 3.1 mJ cm$^-2$), 5.3 ps (3.1 mJ cm$^-2$), 3.8 ps (4.1 mJ cm$^-2$), 3.6 ps (4.6 mJ cm$^-2$).

\begin{figure}
\centering
\includegraphics[width=0.9\linewidth]{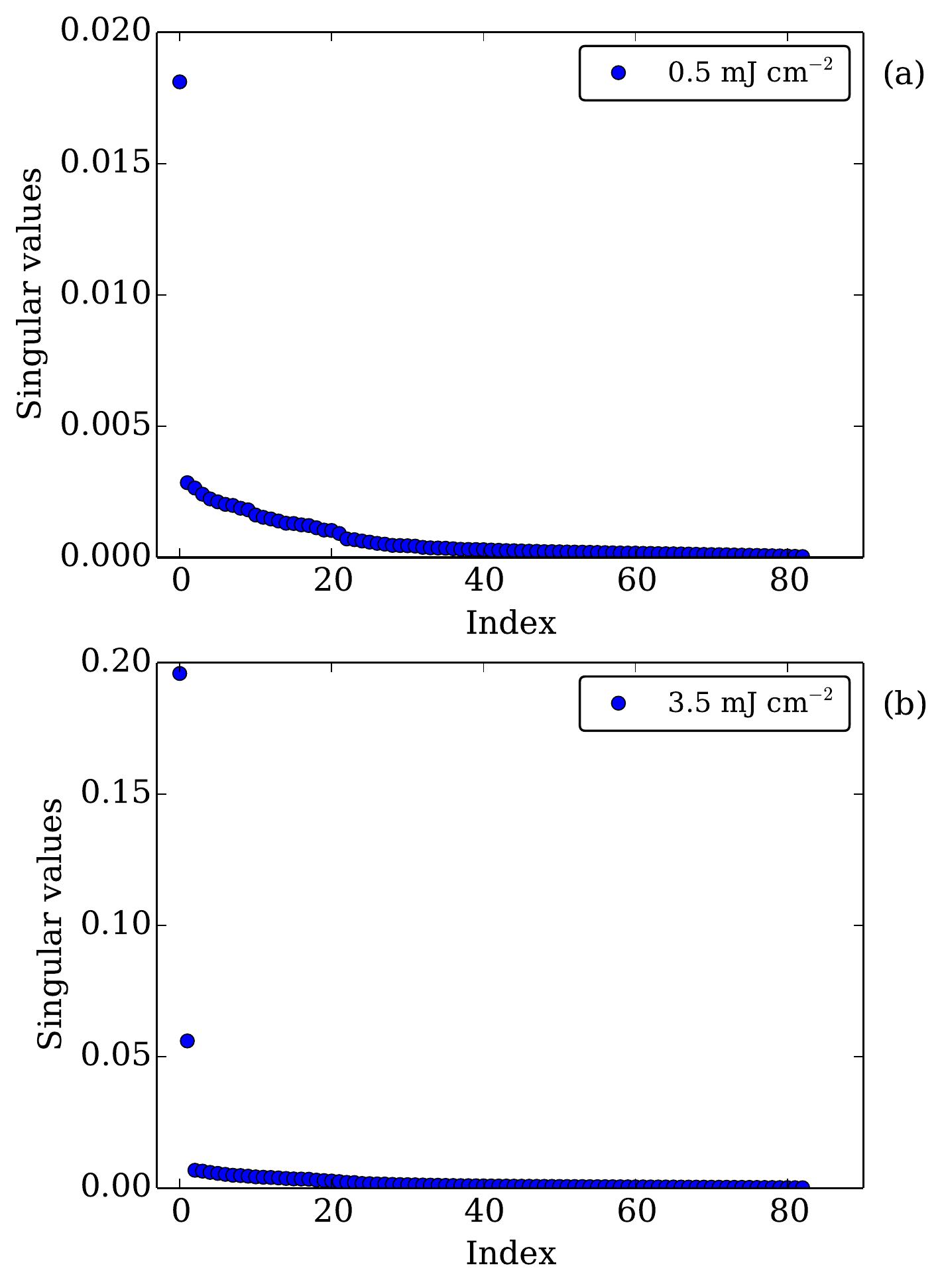}
\caption{Ordered singular values obtained from the singular value decomposition of $[\frac{\Delta R}{R}]_{h\nu,t}$ at 0.5 mJ cm$^{-2}$ (a) and 3.5 mJ cm$^{-2}$ (b).
}
\label{fig:S35B}
\end{figure}

\paragraph{Singular value decomposition} In figure~\ref{fig:S35B} we plot two examples of the result of the singular value decomposition on $[\frac{\Delta R}{R}]_{h\nu,t}$. In the low fluence regime only the largest singular value emerges from the noise (index 0). The variation of the reflectivity can be described as a single spectro-temporal feature and the sample is therefore homogeneous. In the intermediate fluence regime, instead, the singular values relevant with respect to noise are clearly two (indexes 0 and 1). The variation of the reflectivity cannot be described as a single spectral feature evolving in time and phase separation occurs in the sample.\\

\subsection{Results at 80 K}
\begin{figure}[H]
\centering
\includegraphics[width=0.9\linewidth]{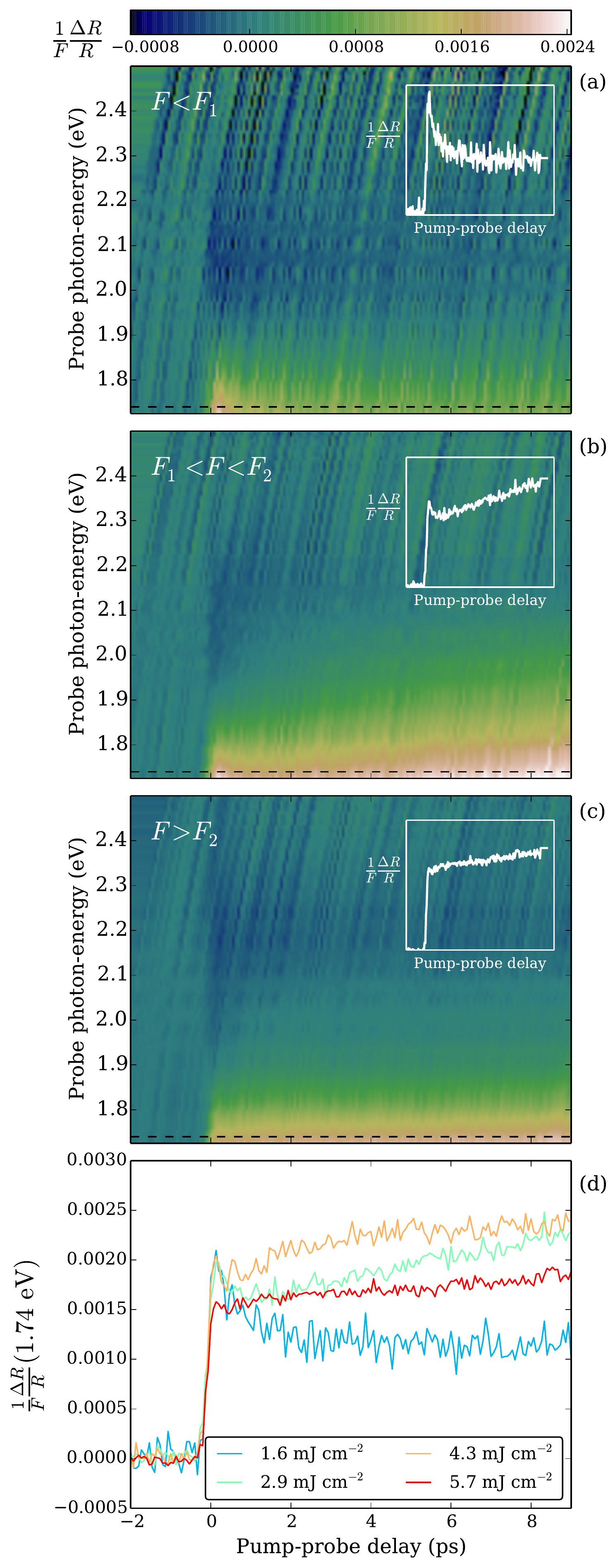}
\caption{(a,b,c): Normalized relative variation of the reflectivity $\frac{1}{F}\frac{\Delta R}{R}$ measured at 80 K as a function of pump-probe delay and probe photon-energy for fluences of 0.6, 2.9, and 5.7 mJ cm$^{-2}$ respectively. (a,b,c) inset: temporal profile of the colour plots at 1.74 eV. Dashed lines: photon-energy corresponding to the insets. (d): $\frac{1}{F}\frac{\Delta R}{R}$(1.74 eV) measured at 80 K for different pump fluences.
}
\label{fig:S80A}
\end{figure}

As mentioned in the main text, the data at 80 K are completely analogous to the ones at 35 K, which have been described in detail. The only difference between the two datasets are the values of the characteristic fluences. In agreement with the thermal interpretation of the energy delivered to the sample by the pump pulses, they are lower at 80 K, since the system needs less energy to reach the transition temperature. In figures~\ref{fig:S80A} and~\ref{fig:S80C} we plot the data measured at 80 K and their analysis, as discussed in the main text for 35 K.

\begin{figure}[H]
\centering
\includegraphics[width=0.9\linewidth]{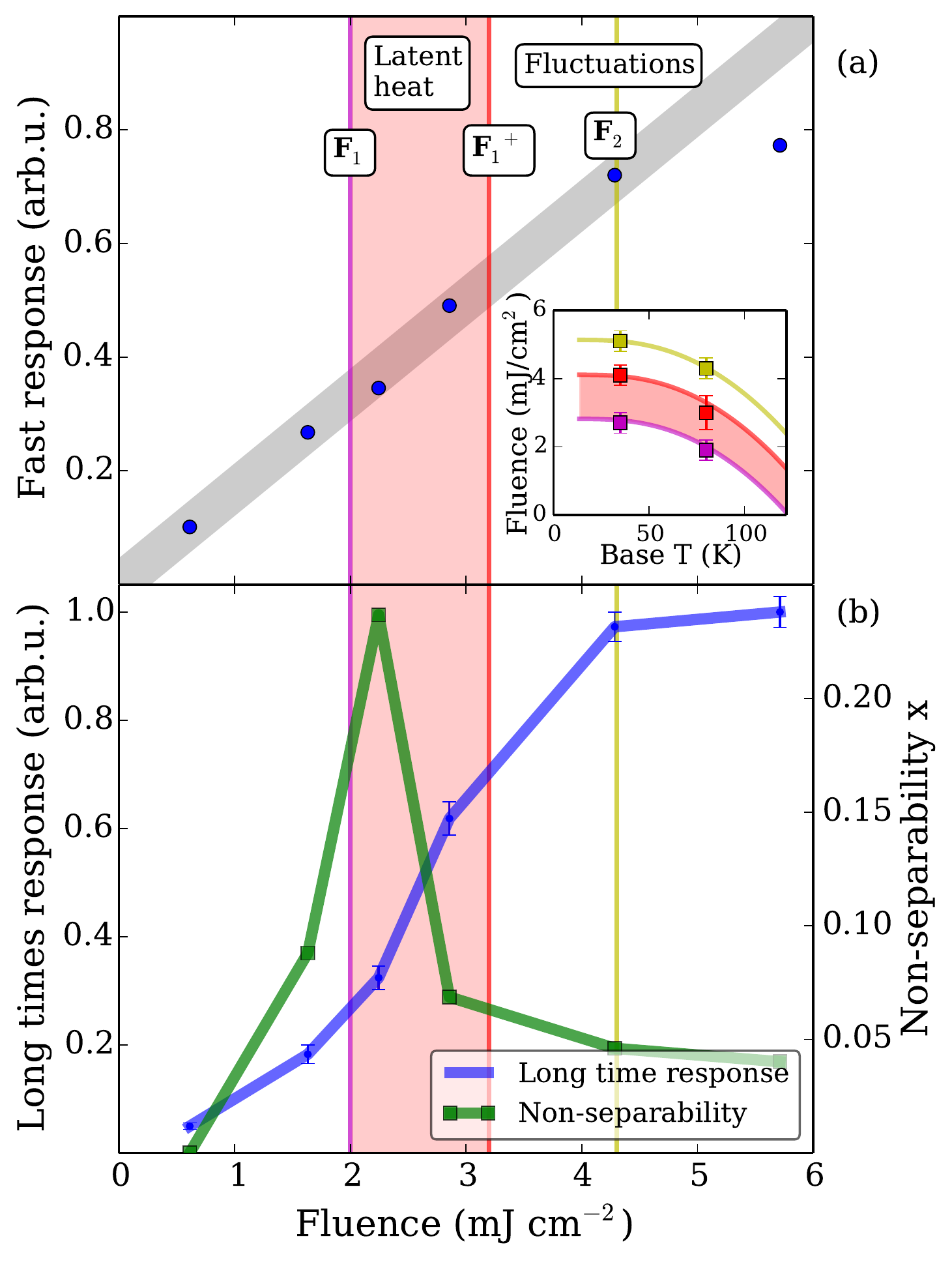}
\caption{(a) ``Fast response'': $\frac{\Delta R}{R}$ at 1.74 eV photon-energy and 0.2 ps pump-probe delay.
(b) ``Long-time response'' (blue curve): $\frac{\Delta R}{R}$ at 1.74 eV photon-energy and 8 ps pump-probe delay. ``Non-separability'' (green curve): ratio of the second largest and largest singular values.
(a inset) The squares are the characteristic fluences extracted from the out-of-equilibrium data as a function of the sample's temperature, corresponding to: onset of the non-linear response (magenta), beginning of the decreasing of the ``non-separability'' (red), saturation of the non-linear response (yellow). The lines represent equivalent fluences calculated from thermodynamic data needed to: reach T$_V^-$ (magenta), reach T$_V^+$ (red), reach 140 K (yellow). The red shaded area corresponds to fluences bringing the sample to T$_V^-$ and supplying part of the latent heat.
Vertical lines and shaded areas in the main figure mimic the inset.
}
\label{fig:S80C}
\end{figure}

\begin{figure}[H]
\centering
\includegraphics[width=0.9\linewidth]{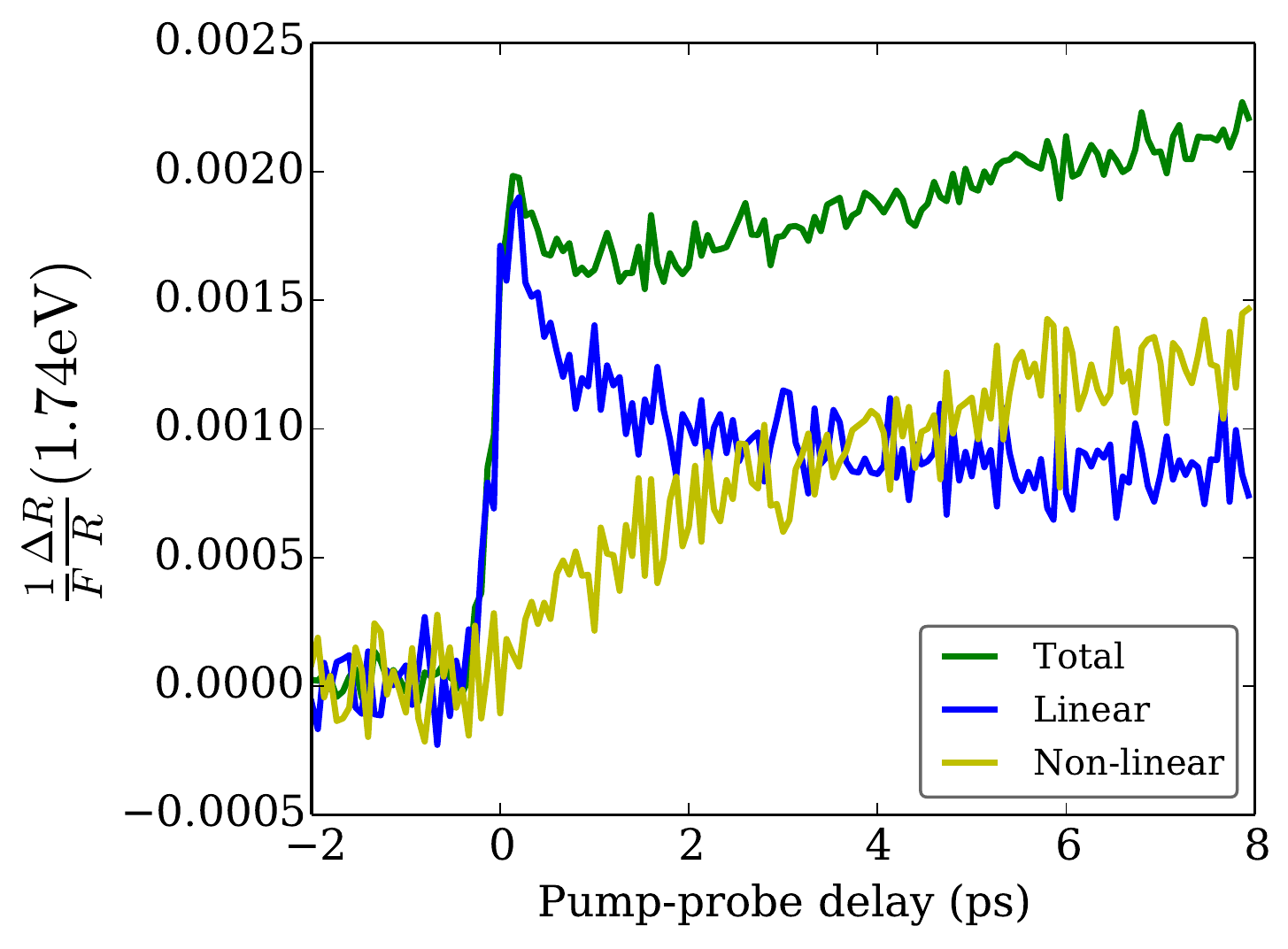}
\caption{Green: $\frac{1}{F}\frac{\Delta R}{R}(t)$ at 1.74 eV for 2.8 mJ cm$^{2}$.  Blue (yellow): linear (non-linear) term of $\frac{1}{F}\frac{\Delta R}{R}$ for F = 2.8 mJ cm$^{-2}$ at 1.74 eV photon-energy.
}
\label{fig:S80C}
\end{figure}

\subsection{Results at 140 K}
\begin{figure}
\centering
\includegraphics[width=0.9\linewidth]{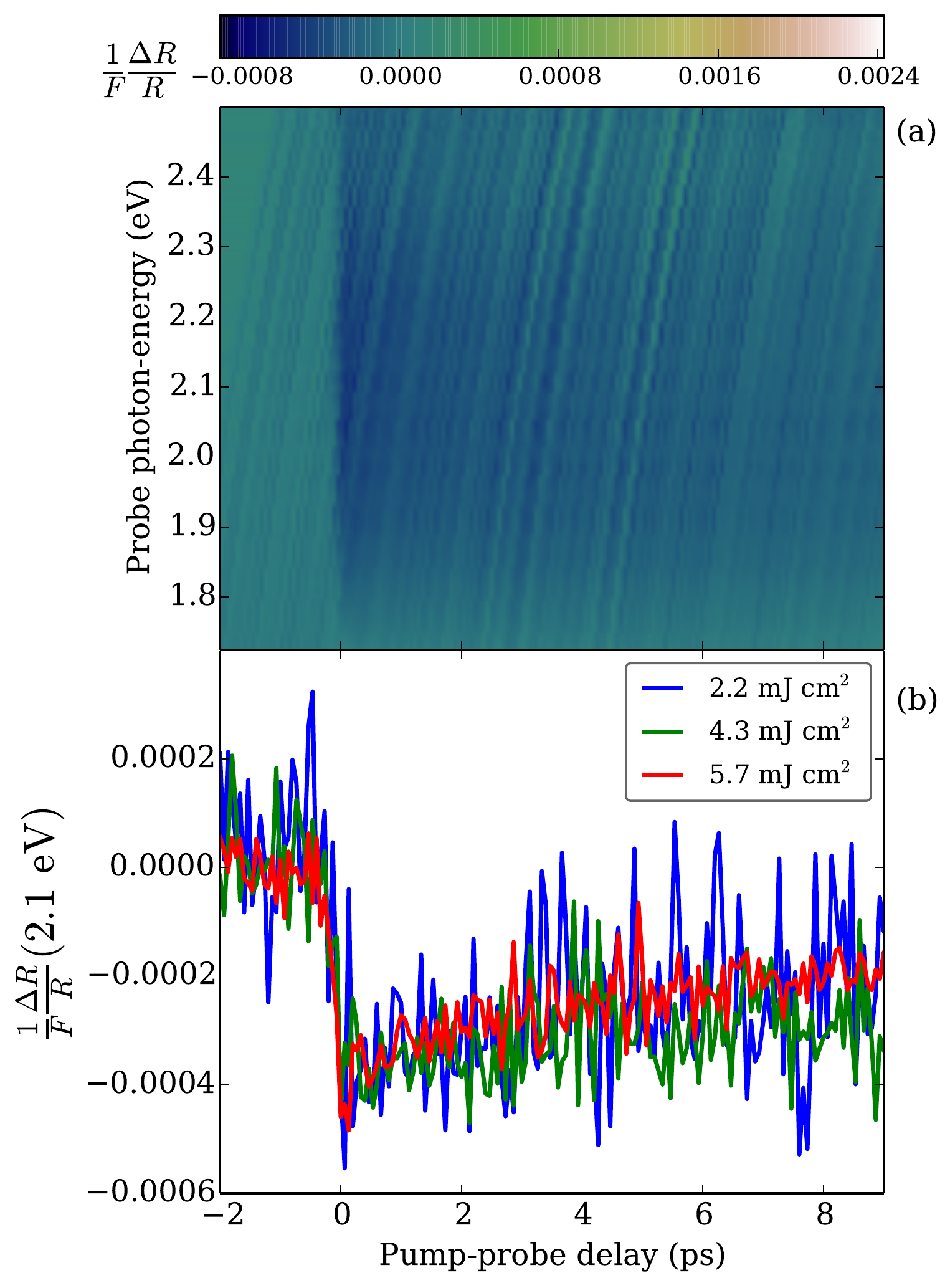}
\caption{ (a): Relative variation of the reflectivity $\frac{\Delta R}{R}$ at 140 K with 5.7 mJ cm$^{-2}$ as a function of pump-probe delay and probe photon-energy. (b) $\frac{\Delta R}{R}$(2.1 eV) at 140 K for three different pump-fluences.
}
\label{fig:S140A}
\end{figure}

In figure~\ref{fig:S140A} we plot $\frac{\Delta R}{R}$ measured at 140 K. In agreement with the interpretation of a photo-induced phase transition for the data at 35 and 80 K, these data display a completely different behaviour. In fact, the spectrum of $\frac{\Delta R}{R}|_{\text{140 K}}$ is different from $\frac{\Delta R}{R}|_{\text{35,80K}}$ in all fluence regimes. Moreover, $\frac{\Delta R}{R}|_{\text{140 K}}$ does not show any non-linearity or slow dynamics arising as a function of fluence. As already mentioned in the main text, also the singular value decomposition points towards an excitation that leaves the sample homogeneous at all the explored pump fluences.

\subsection{Fits of the out-of-equilibrium optical properties}
We performed the fits of $\frac{\Delta R}{R}(h\nu, t)$ in the low fluence regime (figure 8b of the main text) imposing the conservation of the total spectral weight. This constraint gives a more stable result of the fitting procedure. Without it, the spectral weight is approximately conserved in the fit and the result is very similar (both qualitatively and quantitatively) but more noisy. Upon reducing the noise, the constraint enhances the correlation between the two out-of-equilibrium spectral weights (of the 0.6 and 2.0 eV oscillators). In fact the noise of the two curves in figure 8b of the main text, due to closeby local minima in parameter space, is evidently correlated. However, the amplitudes of the variations of the spectral weights and their signs are independent of the details of the fitting procedure.

\bibliography{supplementary}